\documentclass[usenatbib,useAMS,a4paper,fleqn]{mnras}
\usepackage{amssymb}
\usepackage{newtxtext,newtxmath}
\usepackage[T1]{fontenc}
\usepackage{ae,aecompl}
\usepackage{times}

\usepackage{graphicx}
\date{Accepted XXX. Received YYY; in original form ZZZ}
\pubyear{2018}

\begin{document}

\label{firstpage}
\pagerange{\pageref{firstpage}--\pageref{lastpage}}

\title[Unstable low-mass planets polluting WDs]
  {Unstable low-mass planetary systems as drivers of white dwarf pollution}

\author[Mustill et al.]
       {Alexander J.\ Mustill$^1$\thanks{E-mail: alex@astro.lu.se},
         Eva Villaver$^2$,
         Dimitri Veras$^{3,4}$\thanks{STFC Ernest Rutherford Fellow},
         Boris T.\ G\"ansicke$^{3,4}$,
         \newauthor and Amy Bonsor$^5$\\
         $^1$Lund Observatory, Department of Astronomy \& Theoretical Physics,
         Lund University, Box 43, SE-221 00 Lund, Sweden\\
         $^2$Universidad Aut\'{o}noma de Madrid, Departamento de F\'{i}sica Te\'{o}rica, 28049 Madrid, Spain\\
         $^3$Department of Physics, University of Warwick, Coventry CV4 7AL, UK\\
         $^4$Centre for Exoplanets and Habitability, University of Warwick, Coventry, CV4 7AL, UK\\
         $^5$Institute of Astronomy, University of Cambridge, Madingley Road, Cambridge CB3 0HA}

\maketitle

\begin{abstract}
  At least $25\%$ of white dwarfs show atmospheric 
  pollution by metals, sometimes accompanied by detectable circumstellar 
  dust/gas discs or (in the case of WD~1145+017) transiting disintegrating asteroids. 
  Delivery of planetesimals to the white dwarf by orbiting 
  planets is a leading candidate to explain these phenomena. 
  Here, we study systems of planets and 
  planetesimals undergoing planet--planet scattering 
  triggered by the star's post-main sequence mass loss, 
  and test whether this can maintain high rates of delivery 
  over the several Gyr that they are observed. 
  We find that low-mass planets (Earth to Neptune mass) are efficient 
  deliverers of material and can maintain the delivery for Gyr. 
  Unstable low-mass planetary systems reproduce the
  observed delayed onset of significant accretion, as well as the
  slow decay in accretion rates at late times.
  Higher-mass planets are less efficient, 
  and the delivery only lasts a relatively brief time before 
  the planetesimal populations are cleared.
  The orbital inclinations of bodies as they cross 
  the white dwarf's Roche limit are roughly isotropic, implying 
  that significant collisional interactions of asteroids, debris streams 
  and discs can be expected. 
  If planet--planet scattering is indeed responsible for the 
  pollution of white dwarfs, many such objects, and their main-sequence 
  progenitors, can be expected to host (currently undetectable) 
  super-Earth planets on orbits of several au and beyond.
\end{abstract}


\section{Introduction}

\label{sec:intro}

\textcolor{white}{XXXremove me if the hyperref bug goes away}

A significant fraction of white dwarfs (WDs) show evidence of 
possessing remnant planetary systems. This evidence comes 
in the form of photospheric metal pollution 
\citep[e.g.,][]{Zuckerman+03,Koester+14}, 
circumstellar discs of dust and gas 
\citep[e.g.,][]{ZuckermanBecklin87,Graham+90,Gaensicke+06,Farihi16}, and 
in the case of WD~1145+017 a photometric light curve interpreted 
as eclipses of the WD by disintegrating asteroids 
\citep{Vanderburg+15,Gaensicke+16,Rappaport+16,Hallakoun+17}. 
Metal-polluted 
WDs are of particular interest for exoplanet science, as they 
yield insight into the elemental compositions of extrasolar 
planets and asteroids: the gravitational settling timescales 
in most WD atmospheres are astronomically short, and so any 
pollution reflects the composition of recently-accreted material 
\citep{Zuckerman+07,Koester09}. In this way, 
the composition of bodies accreted onto WDs has been compared to 
Solar System objects such as chondritic meteorites and the 
bulk Earth \citep[e.g.,][]{Gaensicke+12,JuraYoung14,Xu+14,ZuckermanYoung17}, 
and in rare cases Kuiper Belt Objects \citep{Xu+17}.

The prevailing paradigm attributes such photospheric pollution 
to the accretion of objects originating from orbits of at least 
several au in the planetary system. Prior to becoming a WD, the 
star's asymptotic giant branch (AGB) 
radius reaches values of roughly $1-5$\,au for 
stars of mass $1-5\mathrm{\,M}_\odot$, resulting in the engulfment 
of planets with comparable perihelion distances; the boundary between 
survival and engulfment is determined by the opposing effects  
of orbital decay due to tidal deformation of the star, and 
orbital expansion due to stellar mass loss 
\citep[e.g.,][]{VillaverLivio09,MustillVillaver12,Villaver+14}. Bodies 
engulfed by the star are expected to be destroyed unless they 
are at least of several Jovian masses 
\citep{VillaverLivio07,Nordhaus+10,Staff+16}. 
Furthermore, the pollutant bodies themselves cannot have very small periapsis 
distances when the star is a giant, lest they be engulfed and destroyed 
before the star becomes a WD. Taken together, these statements
imply the presence of not only a reservoir of 
pollutant bodies, but also one or more objects capable of 
gravitationally perturbing these pollutants onto collision trajectories 
with the WD, all on orbits of several au or beyond. These trajectories 
take the bodies either onto a direct collision course with 
the WD or, more likely, cause their tidal disruption once the 
body crosses the Roche limit at around a Solar radius 
\citep[e.g.,][]{Debes+12,Veras+14}. This disruption is then 
followed by the formation of a circumstellar disc, as fragments are 
subjected to collisional or radiative forces 
\citep{Jura08,Veras+15,Brown+17}.

Thus the general narrative is that the outer regions of planetary 
systems survive the AGB, and can then feed material onto the WD 
by dynamical evolution over the WD's lifetime. The importance of 
these ``outer regions'' (beyond a few au), and the 
relatively large mass of the progenitor stars 
($\sim2\mathrm{\,M}_\odot$), present a challenge to modellers,  
because this is a parameter space inaccessible to most planet-hunting 
surveys. However, they also present an opportunity, because by 
identifying system architectures which do or do not lead to 
the observed incidence and rates of planetesimal accretion 
onto WDs, we can constrain the architectures of planetary systems 
in this otherwise inaccessible parameter space.

Changes to the star as it evolves have significant 
consequences for orbiting bodies. 
Notably, stellar mass loss towards the end of the AGB changes 
the planets' dynamics and stability. The increase in the 
planet:star mass ratios following AGB mass loss makes planet--planet 
interactions stronger and can even destabilise previously-stable 
systems. While the outer planets of the Solar System are sufficiently 
widely-spaced that they will remain stable when the Sun is a WD 
\citep{DuncanLissauer98,Veras16SolarSystem}, 
\cite{DebesSigurdsson02} noted that this will not be the 
case for more tightly-packed systems, hypothesising that 
planetary systems, destabilised by mass loss, may be responsible 
for the observed pollution in WDs.

In recent years, several studies have extended and complemented 
the work of \cite{DebesSigurdsson02}, who focused on the stability of 
systems experiencing a toy model of stellar mass loss. Exploiting 
faster computers, \cite{Veras+13} and \cite{Mustill+14} were 
able to integrate two- and three-planet systems respectively 
over 5\,Gyr with stellar mass and radius evolution taken 
from pre-computed stellar models. \cite{Veras+16} extended these studies 
to planets of unequal masses and of lower masses. An alternative 
line of attack was initiated by \cite{Bonsor+11}, who studied the 
destabilisation of particles orbiting close to a single planet on 
a circular orbit. Subsequent work showed that this requires a 
carefully-constructed chain of planets to feed the material all the 
way to the central object \citep{Bonsor+12}, but \cite{FrewenHansen14} 
showed that a \emph{low-mass, eccentric} planet embedded in a 
planetesimal disc can efficiently feed material to the star; 
\cite{AntoniadouVeras16} showed that a planet on a circular 
orbit cannot. 
\cite{Debes+12} studied the broadening of Kirkwood gaps 
in Asteroid Belt analogues, finding that typical belt masses 
had to be several hundred times greater than our own Asteroid 
Belt to provide the observed accretion rates. 
Recently, \cite{Payne+16,Payne+17} have investigated moons 
liberated from their planets during 
planet--planet scattering as a source of pollution. While 
most papers (including this present one) focus on the planetary 
systems of single stars, a number have studied the effects 
of perturbations from binary companions 
\citep{BonsorVeras15,HamersPortegiesZwart16,PetrovichMuñoz16,Stephan+17} 
as potential drivers of WD pollution. The rich dynamics 
of post-Main Sequence planetary systems were recently 
reviewed by \cite{Veras16}.

The origin of the planetesimals accreted onto the WD can be 
constrained by a spectroscopic determination of the 
composition of the pollutant material.
This methodology is fully equivalent to the analysis of meteorites accreted
onto the Earth, from which fundamental information about the composition of
the Solar System is inferred. Detailed abundance studies of multiple
elements for $\sim20$ debris-polluted white dwarfs
\citep{Zuckerman+07,Klein+10,Gaensicke+12,JuraYoung14,Xu+14,Farihi+16}
and $\sim300$ additional systems in which a handful of elements were detected
\citep{JuraXu13,Hollands+17}
demonstrate that the disrupted exo-planetesimals show a wide range of bulk
compositions, but overall resemble rocky bodies within the inner Solar System,
with one exception being the Kuiper-belt analogue of
\cite{Xu+17}. In addition, a small number of systems
have been found to actively accrete water-rich planetesimals
\citep{Farihi+13,Raddi+15},
and there is further evidence of a larger population of water-bearing planetary
bodies orbiting white dwarfs \citep{Gentile-Fusillo+17},
which highlights the general potential of water delivery to dry inner planets.
On the whole, however, the accreted material resembles asteroids, 
constraining either the place of origin of the accreted planetesimals, 
or their evolution under the large luminosities of 
Red Giant Branch (RGB) and AGB stars.

In this Paper, we explain WD accretion rates by evaluating 
the efficiency of planet--planet scattering in the presence 
of planetesimal discs, synthesising and building on 
a number of previous works. The key ingredients are: 
\begin{enumerate}
\item The finding by \cite{FrewenHansen14} that 
  low-mass eccentric planets can efficiently feed material 
  to a WD over long timescales. This paper did not, however, 
  explain how the required configuration of a single low-mass planet 
  with a high eccentricity ($e\sim0.6$), embedded in a disc 
  of low-eccentricity planetesimals, could attain its initial 
  configuration.
\item The result from \cite{VerasGaensicke15} and 
  \cite{Veras+16} that instabilities in 
  systems of wide-orbit, low-mass planets orbiting WDs can 
  take Gyr to resolve, during which time planets ``meander'', 
  spending significant periods of time on eccentric orbits 
  before ejection occurs. We posit that these instabilities
  offer a natural way for the eccentric planets studied by 
  \cite{FrewenHansen14} to attain the required configuration.
\item The observation from both transit surveys 
  \citep{Fressin+13} 
  and radial-velocity surveys \citep{Cumming+08,Mayor+11} that lower-mass 
  planets are more common than giant planets, 
  with several 10s per cent.\ of stars 
  hosting at least one super-Earth or Neptune within 
  $\sim0.5$\,au. Microlensing reveals a comparable 
  occurrence rate of planets on few au orbits, albeit 
  around lower-mass stars \citep{Shvartzvald+16}. 
  While extrapolation can be dangerous, 
  it is reasonable to speculate that many low-mass 
  planets exist on orbits of a few au and beyond. 
\item The use of a modified \textsc{Mercury} N-body 
  integrator, which allows us to follow accurately 
  the evolution of planetary systems 
  including test particles for several Gyr. Our integrator 
  is based on that described by 
  \citet[where we adapted the Bulirsch--Stoer integrator]{Veras+13}, but 
  has been modified to make use of the more accurate RADAU 
  algorithm: in Mustill et al.\ (in prep) we show that this accurate integrator 
  is essential to accurately determine instability timescales 
  for the systems we consider.
\end{enumerate}

\begin{figure*}
   \includegraphics[width=\textwidth]{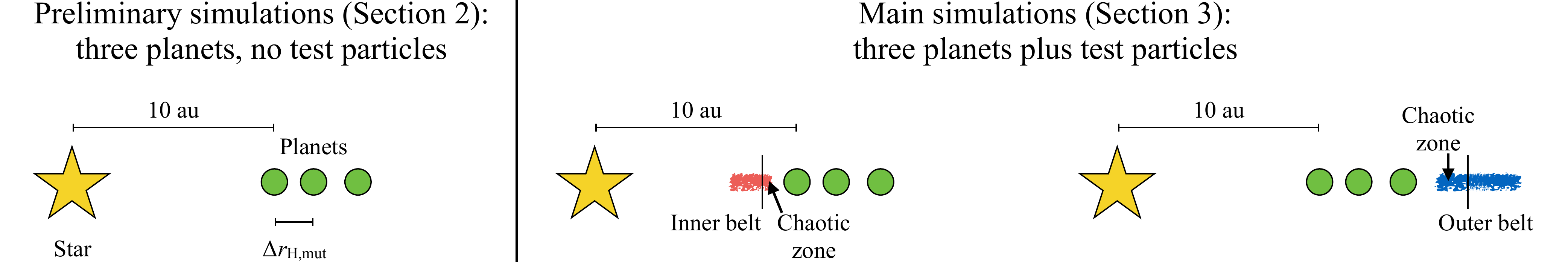}
   \caption{Schematic setup for our simulations. In Section~\ref{sec:notp}
     we consider 3-planet systems orbiting a star of initial mass 
     $3\mathrm{\,M_\odot}$. In Section~\ref{sec:tp} we incude 
     test particles in either an inner belt or an outer belt, in 
     each case extending from within the nearest planets chaotic 
     zone to more distant stable orbits.}
   \label{fig:setup}
 \end{figure*}

We first conduct simulations of unequal-mass three-planet 
systems over host stars' MS and WD lifetimes (for 5\,Gyr), to 
explore the range of stability times for planetary systems of different 
masses (Section~\ref{sec:notp}). We then run a smaller 
number of integrations of three-planet systems incorporating discs 
of test particles, to study the delivery rate of these bodies 
to the WD (Section~\ref{sec:tp}). We study both ``inner belts'' 
of asteroidal bodies interior to the planets' orbits, 
and ``outer belts'' of cometary bodies exterior to the planets' orbits, 
to compare the relative efficiencies of delivery of the 
two populations. A cartoon showing the system configurations we 
study is shown in Figure~\ref{fig:setup}. We then discuss these results 
 in the context of the latest observations of WD pollution 
 (Section~\ref{sec:discuss}), and conclude in Section~\ref{sec:conclude}.

 \section{Preliminary integrations: no test particles}

 \label{sec:notp}

 \subsection{Setup}

 We run several sets of integrations of three-planet systems with 
 no test particles in order to identify architectures of interest 
 for more detailed study. These integrations themselves extend the 
 range of system architectures covered in previous studies: where 
 \cite{DebesSigurdsson02} and \cite{Mustill+14} both studied 
 equal-mass systems, \cite{Veras+16} studied unequal-mass systems 
 based on permutations of the Solar System planets. Here we 
 extend the variety of unequal-mass systems, considering 
 three mass-ranges: roughly Saturn to super-Jupiter (S2SJ), Neptune 
 to Saturn (N2S), and super-Earth to Neptune (SE2N).

 We take a $3\mathrm{\,M}_\odot$ star for all our simulations. This 
 is at the upper range of the progenitor masses of polluted 
 white dwarfs \citep{Koester+14}, but 
 allows our simulations to be conducted in 
 a reasonable time (recall the very strong dependence 
 of stellar main-sequence lifetime on initial mass: a $3\mathrm{\,M}_\odot$ 
 star will live for $\sim0.5$\,Gyr but a $1\mathrm{\,M}_\odot$ star 
 for $\sim12$\,Gyr). 
 The star is evolved with SSE \citep{Hurley+00}, and the 
 output file containing mass and radius is fed to the 
 modified \textsc{Mercury} integrator 
 \citep[][Mustill et al in prep; and 
     see Appendix~\ref{sec:integrator}]{Veras+13}. 
 The star becomes a WD of mass $0.75\mathrm{\,M}_\odot$ at 
 $477.6$\,Myr, after attaining a peak AGB radius of $2.85$\,au. 
 Stellar mass loss is implemented in 
 \textsc{Mercury} by adjusting the stellar 
 mass both at major integrator timesteps and at all 
 sub-intervals of these timesteps \citep{Veras+13}. 
 Mass loss is assumed to be isotropic 
 \citep[a good approximation for bodies at the orbital 
   radii we consider,][]{VerasHT13}, but we naturally 
 capture any non-adiabatic effects from rapid mass 
 loss\footnote{In the adiabatic limit where the mass loss 
 timescale is much longer than the orbital timescale, 
 planets' orbits simply expand at constant eccentricity 
 to conserve angular momentum. Rapid mass loss can 
 excite bodies' eccentricities. See \cite{Hadjidemetriou63} 
 and \cite{Veras+11} for details.}.
 Systems are integrated for 5\,Gyr with the RADAU 
 integrator from the \textsc{Mercury} package 
 \citep{Chambers99}, with a tolerance parameter 
 of $10^{-11}$. Particles are removed from the simulation 
 when they are ejected, or collide with a planet or 
 with the physical radius of the star. We record the pericentres 
 of bodies throughout the integrations, allowing us to 
 post-process the simulation data to check for the intrusion 
 of particles past the Roche limit of the WD at $\sim0.005$\,au.

 As in our previous works, we set the inner of 
 the three planet's initial semi-major
 axis to $10$\,au. This allows faster integrations (thanks to the larger
 timestep for a given tolerance), but also ensures that the planets
 we consider are well beyond the reach of tidal forces when the
 star is a large AGB object\footnote{For stars of this mass, the AGB
 tip radius of $\sim3$\,au is considerably larger than the maximum RGB
 radius of $\sim0.2$\,au.} 
 \citep{MustillVillaver12,NordhausSpiegel13}. 
 The limit for surviving tidal inspiral is
 $\sim4$\,au for a Jovian planet on a circular orbit around a
 $3\mathrm{\,M}_\odot$ primary, and $\sim2.5$\,au for Earth-mass 
 planets. Planets on slightly wider 
 orbits will still experience some tidal orbital decay 
 even though they avoid engulfment. While $10$\,au
 is conservative for planets on circular orbits, it also ensures that
 our inner belts of test particles in the full integrations described in the
 next section are also safe from engulfment on the AGB.

 \subsubsection{Generating a planet population}

 \label{sec:planet_setup}

 Unfortunately, the properties of planet populations 
 (most importantly for our purposes, mass 
 and semimajor axis) are almost entirely unknown 
 around both WDs and their largely intermediate-mass progenitors. 
 Direct imaging is sensitive to very massive super-Jupiter planets 
 at 10s to 100s of au, but with the exception of the extremely 
 wide-orbit ($\sim2500$\,au) companion to WD~0806-661 
 \citep{Luhman+11} there have been 
 no detections of planetary companions to single WDs 
 \citep{Debes+05a,Debes+05b,Burleigh+08,
   ZinneckerKitsionas08,Hogan+09,Xu+15}. 
 Searches for IR excess are sensitive 
 to closer, unresolved, companions, but have similarly given 
 negative results \citep{Farihi+08}. Timing of stellar pulsations, 
 sensitive to giant planets at a few au, has also given no robust 
 detections \citep{Mullally+08,Winget+15}. Finally, there have been 
 no detections of transiting planets (as opposed to asteroids) 
 orbiting white dwarfs on small orbits $\lesssim0.1$\,au 
 \citep{Faedi+11,Fulton+14,Sandhaus+16,VanSluijsVanEylen17}.
 Currently, the strongest limits come from Pan-STARRS1 for giant 
 planets \citep[$\lesssim0.1\%$ for Jovian planets at 
   $0.01$\,au;][]{Fulton+14} and from \emph{K2} for smaller 
 planets \citep[$\lesssim25\%$ for Earth-radius planets 
   at $0.01$\,au;][]{VanSluijsVanEylen17}, although 
 occurrence rates beyond $0.05$\,au are weak to non-existent.

 The progenitors of white dwarfs may be probed for planets 
 with radial velocity surveys when they have evolved to leave the 
 main sequence. \cite{Reffert+15} found 
 that the occurrence rate of Jovian planets peaks for stellar masses 
 around $2\mathrm{\,M}_\odot$, at $5-15\%$ (depending on whether one counts 
 only the secure detections or includes candidates) on periods of up to a few years. 
 Direct imaging surveys have probed orbits of several 10s of au 
 for super-Jovian planets but, despite spectacular finds such as 
 the four planets of HR~8799 orbiting a $1.5\mathrm{\,M}_\odot$ primary 
 \citep{Marois+08,Marois+10}, the occurrence rate of such systems is low 
 \citep[$\lesssim10\%$,][]{Nielsen+13,Chauvin+15}. Unfortunately, 
 less massive planets remain undetectable, unless the object 
 Fomalhaut~b \citep{Kalas+08} is in fact a super-Earth or ice giant 
 surrounded by a dust cloud \citep{KennedyWyatt11}.

 This ignorance means that we must use a set of artificially-constructed
 systems which cover a range of possible architectures for 
 the outer regions of planetary systems orbiting intermediate-mass 
 stars. We can however be guided by the much better-understood 
 populations of planets on smaller ($\lesssim1$\,au) orbits 
 around Solar-type and M~dwarf stars. Radial-velocity surveys 
 have found that, within an orbital period of 100 days, 
 around $50\%$ of FGK stars have at least one detectable planet 
 of mass $m\sin i<30\mathrm{\,M}_\oplus$, while $\sim10\%$ host 
 a gas giant $m\sin i>50\mathrm{\,M}_\oplus$ with a period of 
 up to 10 years \citep{Mayor+11}. These numbers are in agreement 
 with the statistics of transiting planet candidates from the 
 \textit{Kepler} mission, which found that $50\%$ of stars have 
 a planet of radius $0.8-22\mathrm{\,R}_\oplus$ with a period 
 less than 85 days and $5\%$ host a giant planet (radius 
 $>6\mathrm{\,R}_\oplus$) with a period less than 400 days 
 \citep{Fressin+13}. 
 Thus, low-mass planets are common close to the star. 
 It is not unreasonable to speculate that low-mass (super-Earth 
 or Neptune-like) planets may be equally common on wide orbits 
 around the progenitors of white dwarfs, although this depends 
 on the properties of protoplanetary discs and the 
 pathways planet formation takes around these more massive 
 stars.

 A final consideration is the spacing of orbits in multi-planet 
 systems. The eccentricity distribution of giant planets 
 is explained if the majority of them 
 \citep[$\sim80\%$][]{Raymond+11,JuricTremaine08} 
 were sufficiently tightly packed after formation that they experienced 
 scattering early in their main sequence evolution. 
 Systems slightly more widely spaced are prime candidates 
 for destabilisation following post-MS mass loss, 
 while systems that have already experienced one 
 instability on the MS often experience 
 a second after the star becomes a WD \citep{Mustill+14}. 
 The timescale for a system to experience instability 
 can be crudely estimated from the planets' separations 
 in mutual Hill radii \citep{Chambers+96}. 
 For observed \textit{Kepler} systems of four or more 
 planets, \cite{PuWu15} found that the distribution of 
 separations peaks at 14 mutual Hill radii. 
 \cite{Lissauer+11} and \cite{Fabrycky+14} estimated, from the empirical 
 scalings of \cite{SmithLissauer09}, that a separation of 9 
 mutual Hill radii should be the approximate limit for 
 systems to remain stable to the mid-MS age of a typical star, 
 but nevertheless found a number of more tightly-packed systems. 
 Considering all \textit{Kepler} multiples, \cite{Weiss+17} 
 recently found that the median separation is $\sim20$ mutual 
 Hill radii. If wider-orbit planets follow similar spacings, not 
 all are expected to be destabilised by mass loss: \cite{Mustill+14} 
 estimate the maximum limit for systems of 
 three Earth-mass planets to be destabilised 
 around white dwarfs at around 18 single-planet Hill radii, or 14 
 mutual Hill radii.

 Based on the above considerations, we construct the following simulation 
 sets, each of 128 runs for a total of 1\,536 runs: 
 \begin{itemize}
   \item \textsc{s2sj-XYrh:} Three giant planets, masses chosen from 
     $100-1000\mathrm{\,M}_\oplus$ (``Saturn to 
     super-Jupiter''). The innermost planet 
     is placed at 10\,au, and the subsequent planets are separated 
     by $x$ to $y$ mutual Hill radii, where $y=x+2$ and $x\in\{4,5,6,7,8\}$.
   \item \textsc{n2s-57rh:} Three intermediate-mass planets, masses chosen 
     from $10-100\mathrm{\,M}_\oplus$ (``Neptune 
     to Saturn''). We perform a single set of runs in this mass range, 
     with the inner planet again at 10\,au and subsequent planets 
     spaced by $5-7$ mutual Hill radii.
   \item \textsc{se2n-XYrh:} Three low-mass planets, masses chosen from 
     $1-30\mathrm{\,M}_\oplus$ (``Super-Earth to Neptune''), 
     with the inner planet at 10\,au and subsequent planets 
     separated by $x$ to $y$ mutual Hill radii, where $y=x+2$ and 
     $x\in\{5,6,7,8,9,10\}$.
 \end{itemize}
 Extrapolating from the statistics of close-in planets 
 discussed above, we expect the \textsc{se2n} runs to 
 represent the mass range of perhaps the majority of 
 the progenitor systems' planets. However, our systems 
 are rather tightly spaced compared to the average 
 \textit{Kepler} multiple.

 For each planetary system, the masses of the planets are drawn 
 independently from a distribution uniform in the logarithm of the mass, 
 within the specified range. 
 The initial eccentricities are zero, and inclinations are drawn in the 
 range $[0^\circ,1^\circ]$ from a reference plane, with randomised 
 longitudes of ascending node and mean anomalies. The left-hand panel 
 of Figure~\ref{fig:setup} illustrates the setup.

 \subsection{Results}

 \begin{figure}
   \includegraphics[width=0.5\textwidth]{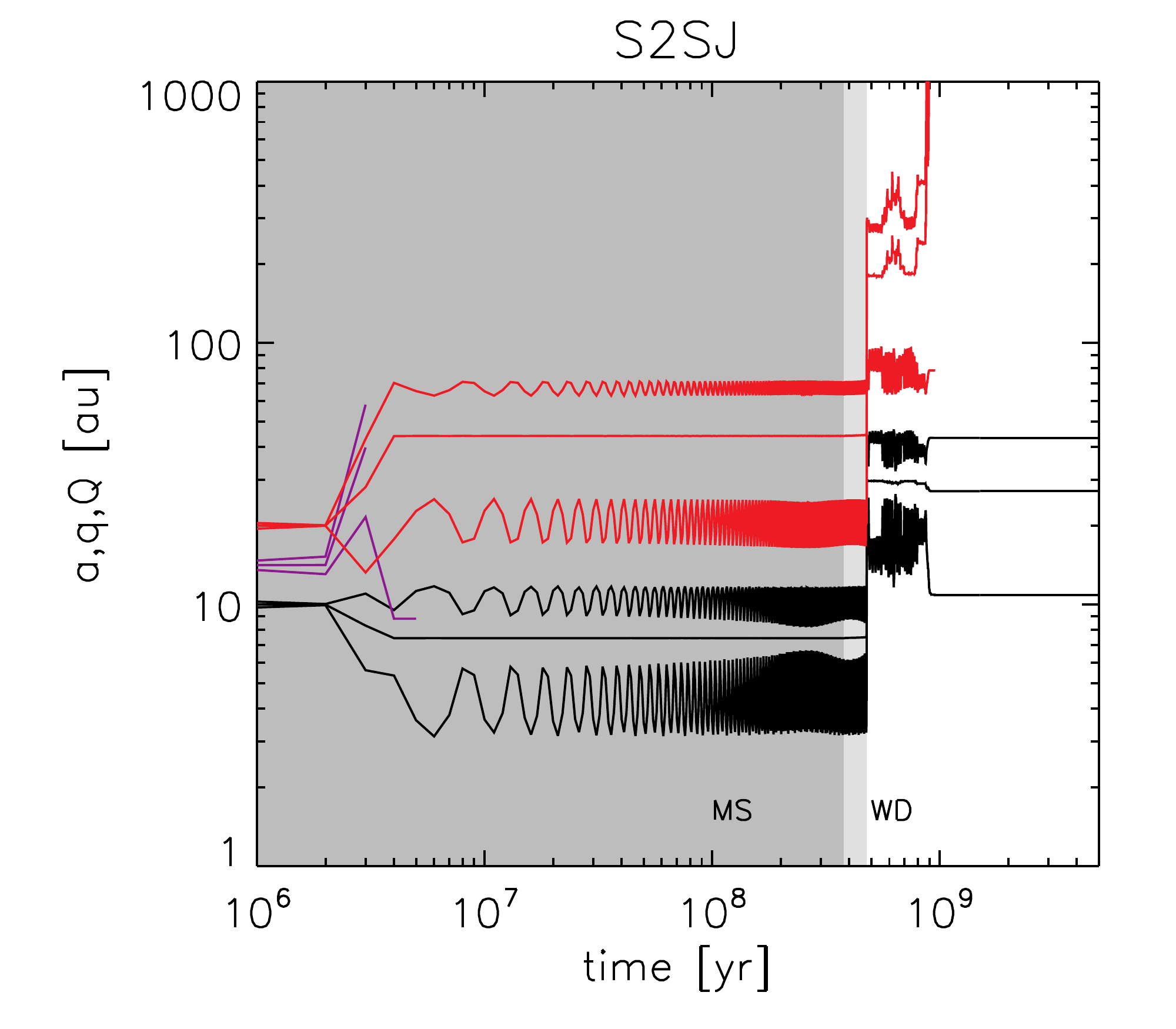}
   \caption{Example evolution of one system of three planets 
     orbiting an evolving star, from 
     simulation set \textsc{s2sj-57rh}. For each planet 
     (shown in a different colour), the semimajor axis $a$, 
     pericentre $q$ and apocentre $Q$ are shown as a function of time. 
     The stellar evolutionary state is shown as the background 
     shading (dark for main sequence, light for end MS to end AGB, 
     and white for white dwarf). 
     The system experiences an early instability, resulting in the ejection of 
     one planet after a few Myr. The resulting two-planet configuration 
     remains stable throught the star's remaining MS lifetime, before 
     mass loss just before $500$\,Myr results in a second instability 
     and eventual ejection of another planet. Planet--planet scattering continues 
     for several hundred Myr before eventual ejection.}
   \label{fig:aqQ}
 \end{figure}

 These preliminary integrations show the usual destabilising 
 effect of stellar mass loss on orbiting planets. Stellar mass loss 
 triggers instability by increasing the planet:star mass ratio, 
 increasing the size of the Hill spheres and broadening 
 orbital resonances. An example 
 is shown in Figure~\ref{fig:aqQ}. This system experiences 
 one early instability on the MS (at around 2\,Myr) 
 before settling into a stable 
 two-planet configuration. This configuration is itself 
 destabilised following mass loss on the AGB, losing a 
 second planet at around $0.9$\,Gyr, following several 
 hundred Myr of planet--planet scattering.

 \begin{figure*}
   \includegraphics[width=0.48\textwidth]{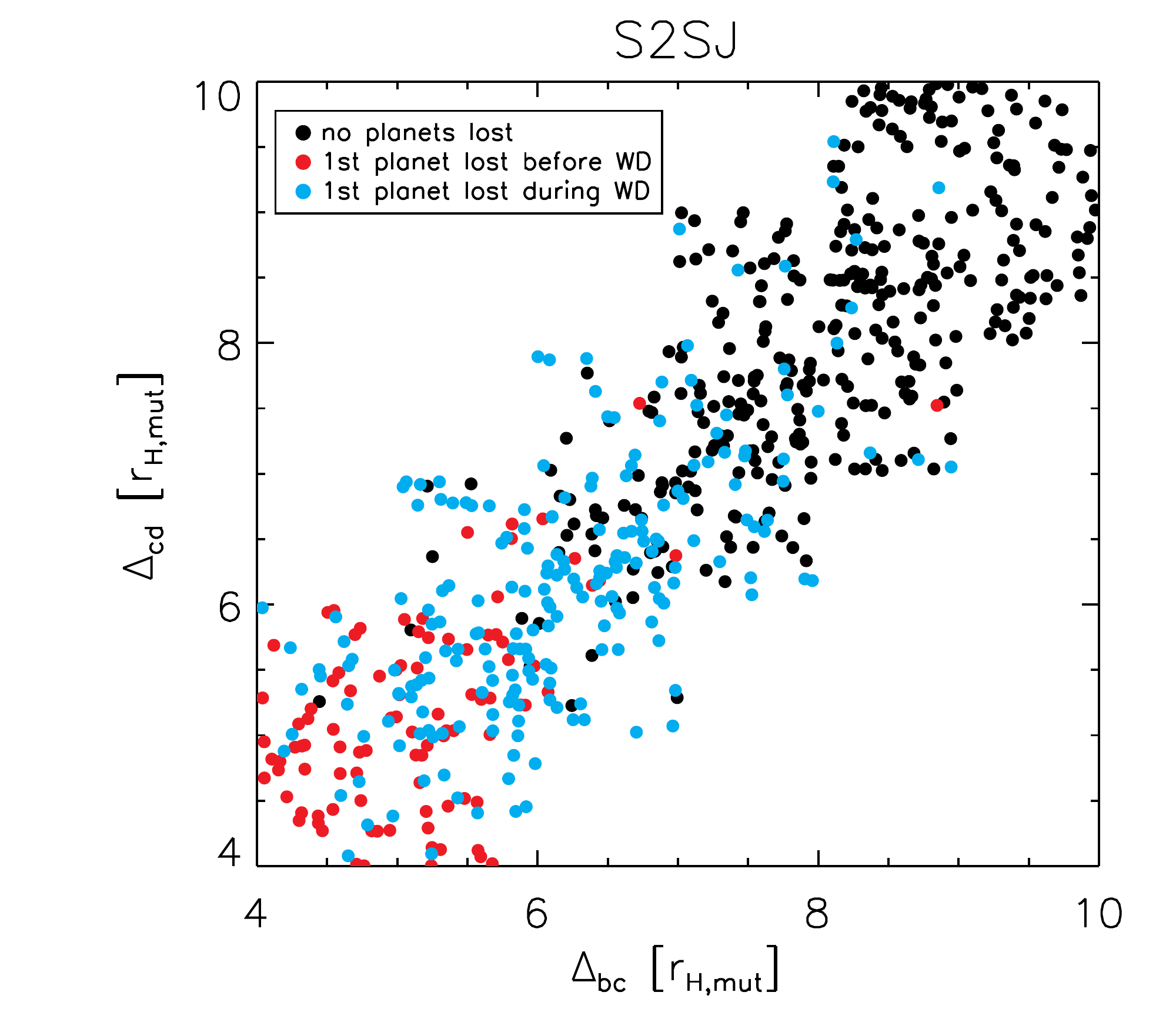}
   \includegraphics[width=0.48\textwidth]{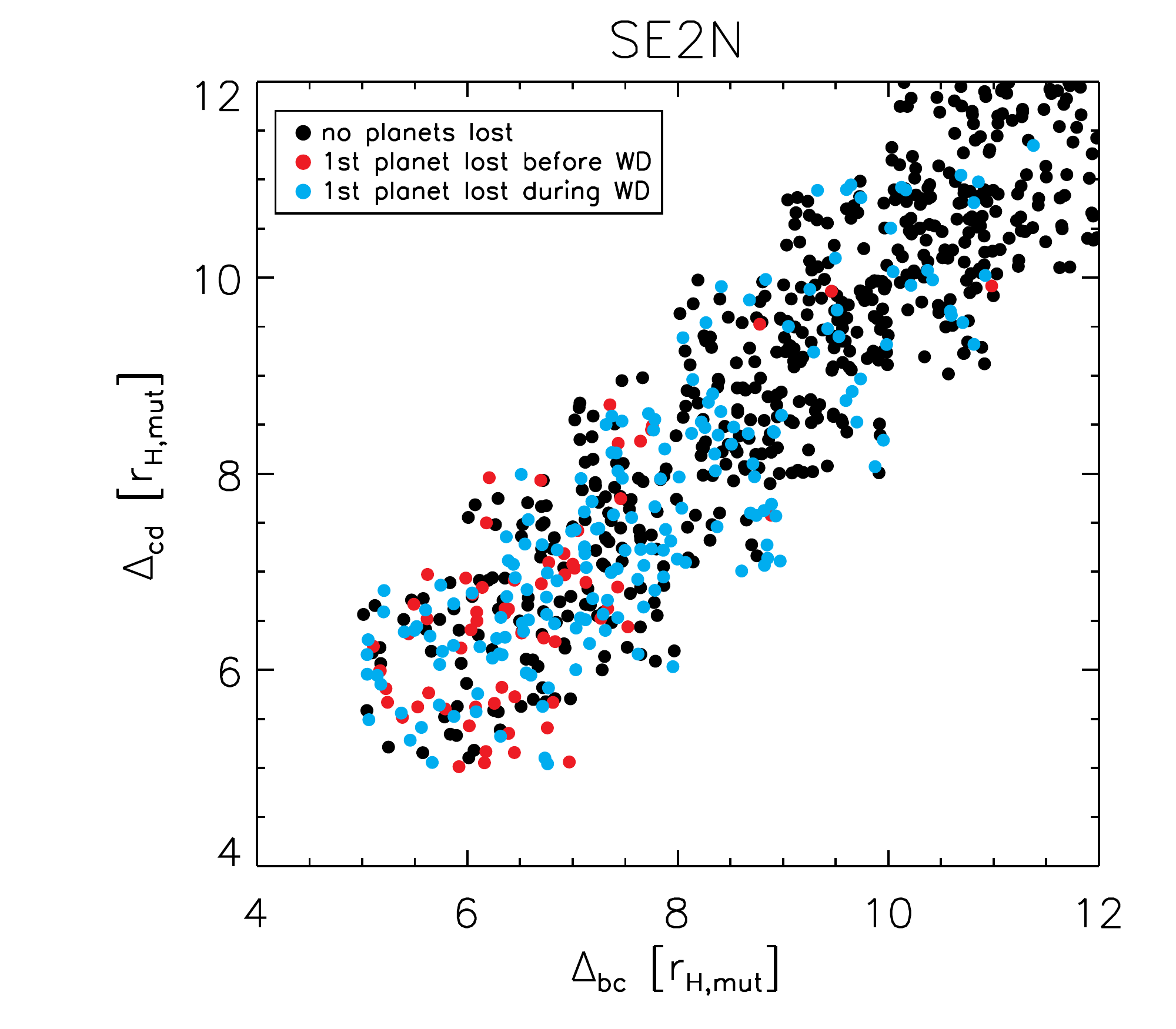}
   \caption{Integration outcomes for systems of three planets, as a function of 
     the initial separation in mutual Hill radii between the inner 
     two planets ($\Delta_\mathrm{bc}$) and the outer two 
     planets ($\Delta_\mathrm{cd}$). \textbf{Left: }
     \textsc{s2sj} runs (planet masses ``Saturn to super-Jupiter''). 
     \textbf{Right: }\textsc{se2n} runs (planet masses ``super-Earth to Neptune''). 
     As the separation is increased, there is a transition from 
     systems unstable on the MS, to those destabilised by mass loss, 
     to those stable throughout the integration, but the 
     transition is not abrupt. We do not show here the \textsc{n2s}
       runs as we only considered a small range in separations, but 
       the same trend is present.}
   \label{fig:delta-delta}
 \end{figure*}

 Planetary instability may occur before or after AGB mass loss, 
 or not at all, the timescale being crudely set by the 
 planetary separation \citep[e.g.,][]{Chambers+96}. We show the 
 stages of evolution at which systems first lost planets, to 
 collision or ejection, as 
 a function of the initial separation in mutual Hill radii, 
 in Figure~\ref{fig:delta-delta}. As the separation increases, there 
 is a trend away from instability before mass loss, to instability 
 following mass loss, and finally to stability for the 
 whole integration. This transition is not abrupt and considerable 
 overlap exists between these regimes, as expected from 
 previous studies \citep[e.g.,][]{Chambers+96,Mustill+14}. 
 The transition is less abrupt for the lower-mass planets. 
 Instability may also be measured by the onset of orbit-crossing, 
 and by this criterion the instabilities at different ages are tabulated 
 in Table~\ref{tab:notp}. Again, we see a trend away from 
 most instability occurring before AGB mass loss, to most 
 instability occurring after AGB mass loss, to finally most systems 
 being stable for the entire 5\,Gyr integration time.

 \begin{table}
   \begin{tabular}{lccc}
     Simulation set       & \# runs & \# orbit-crossing begins & \# orbit-crossing begins\\
                          &         &    before WD forms    &     after WD forms\\
     \hline
     \textsc{S2SJ-46rh}   & 128     &  72                   & 55 \\
     \textsc{S2SJ-57rh}   & 128     &  21                   & 85 \\
     \textsc{S2SJ-68rh}   & 128     &   3                   & 48 \\
     \textsc{S2SJ-79rh}   & 128     &   1                   & 17 \\
     \textsc{S2SJ-810rh}  & 128     &   0                   &  5 \\
     \\
     \textsc{N2S-57rh}    & 128     &  62                   & 63 \\
     \\
     \textsc{SE2N-57rh}   & 128     &  99                   & 29 \\
     \textsc{SE2N-68rh}   & 128     &  49                   & 72 \\
     \textsc{SE2N-79rh}   & 128     &  17                   & 91 \\
     \textsc{SE2N-810rh}  & 128     &   7                   & 67 \\
     \textsc{SE2N-911rh}  & 128     &   3                   & 42 \\
     \textsc{SE2N-1012rh} & 128     &   0                   & 17
   \end{tabular}
   \caption{Instabilities in triple-planet systems without 
     test particles. ``Instability'' here refers to the onset 
     of orbit-crossing or the loss (collision or ejection) of 
     the first planet, if orbit-crossing is not picked up 
     because of the discrete output intervals.}
   \label{tab:notp}
 \end{table}

\begin{figure*}
   \includegraphics[width=.33\textwidth]{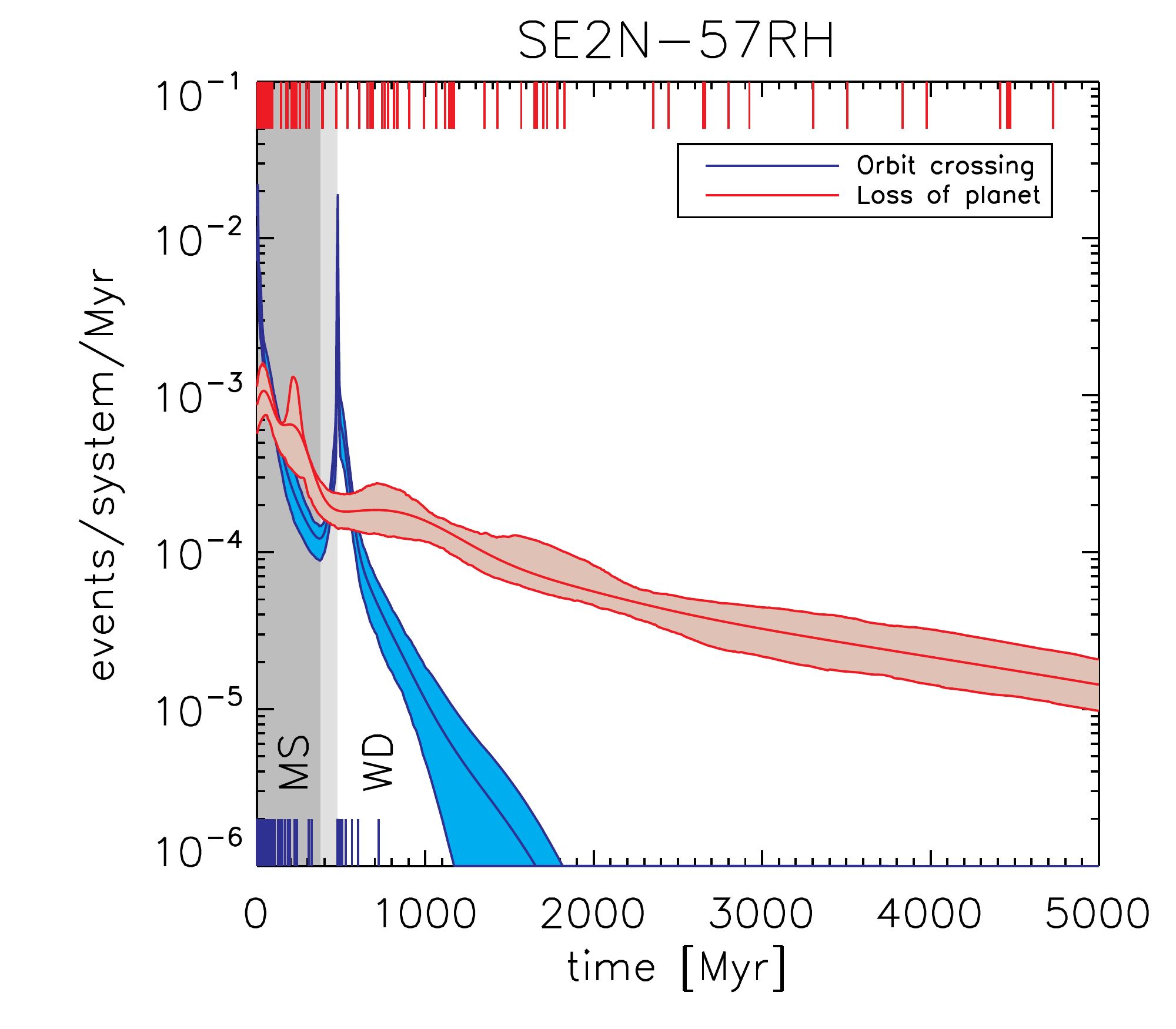}
   \includegraphics[width=.33\textwidth]{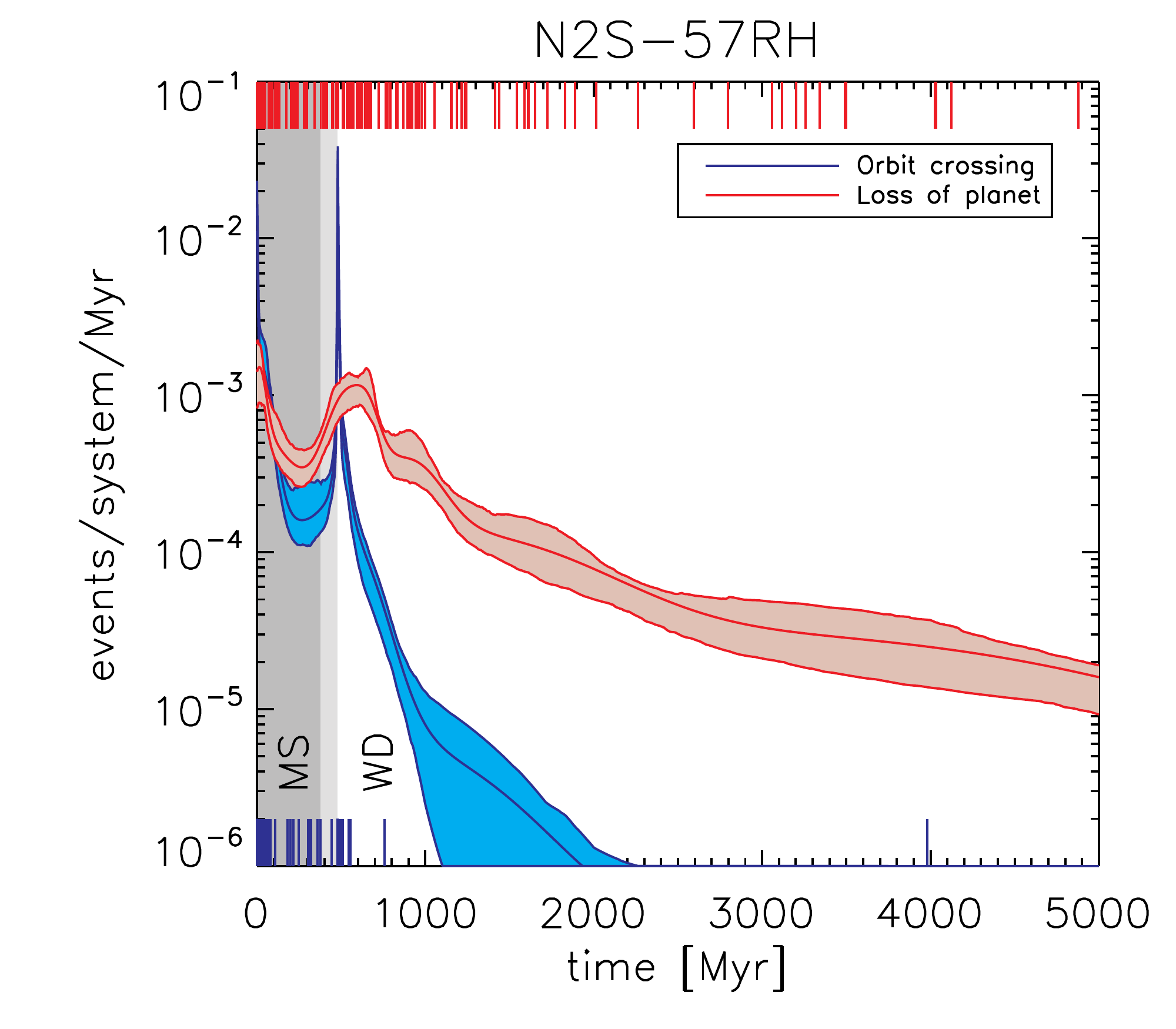}
   \includegraphics[width=.33\textwidth]{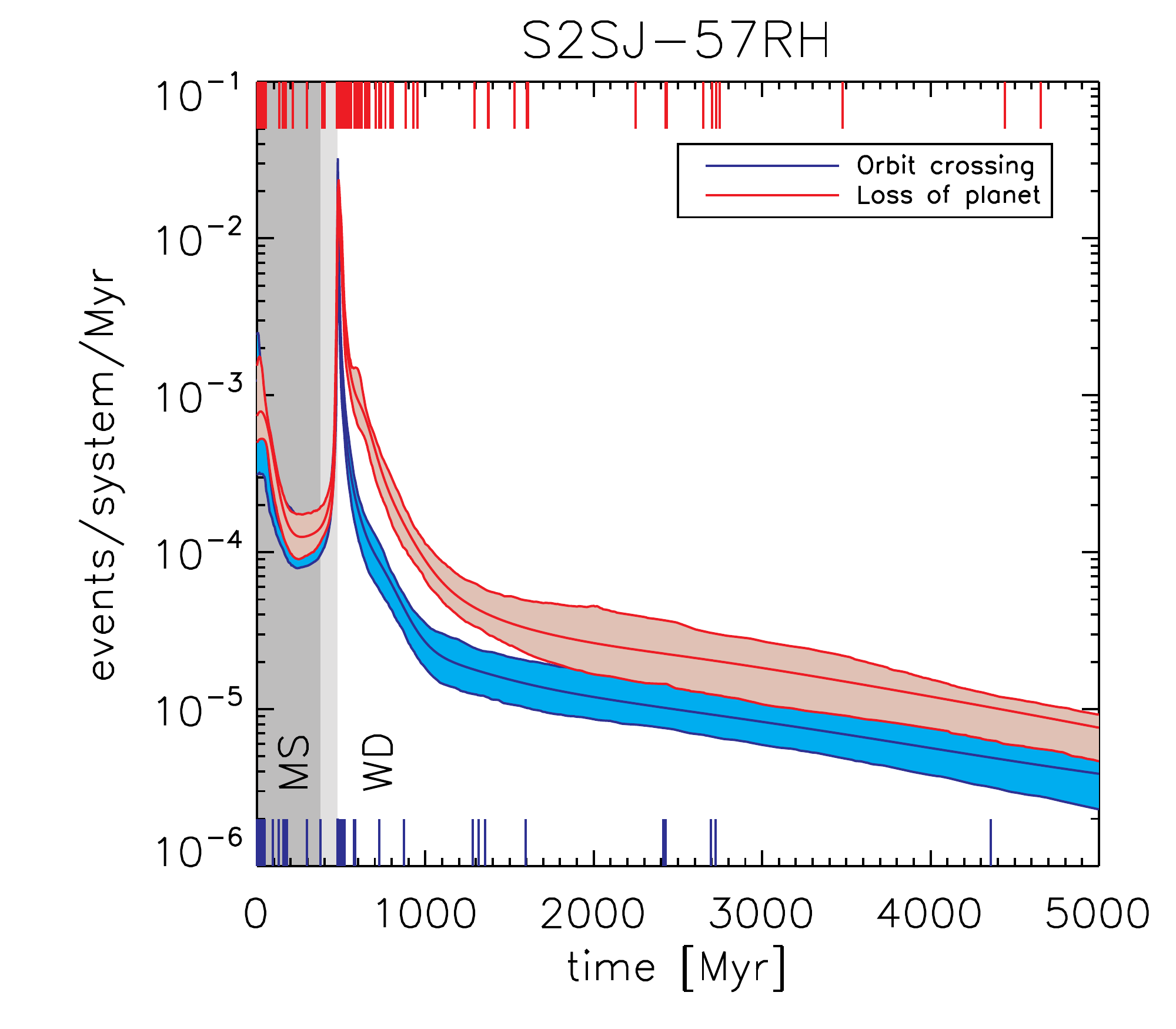}
   \includegraphics[width=.48\textwidth]{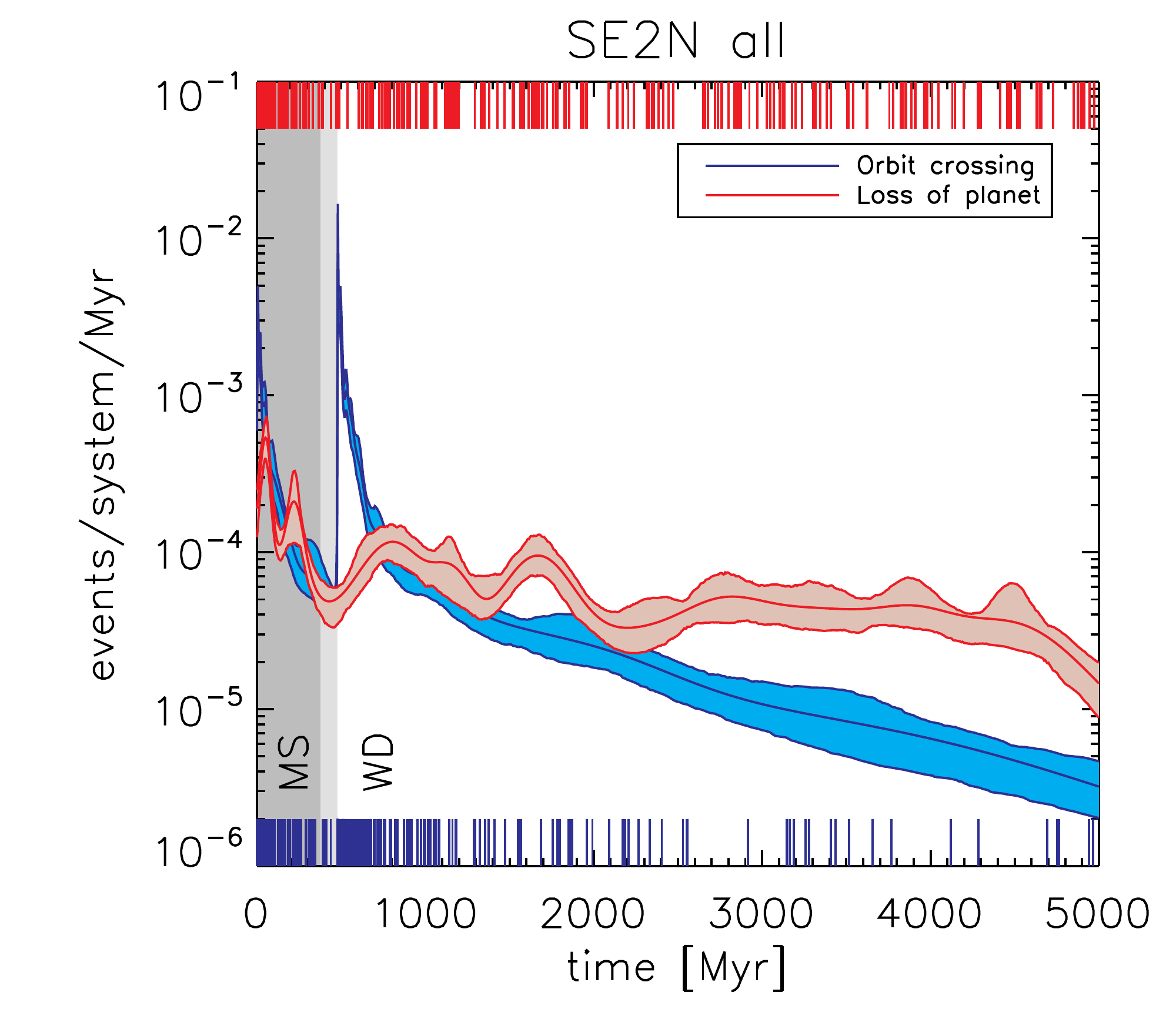}
   \includegraphics[width=.48\textwidth]{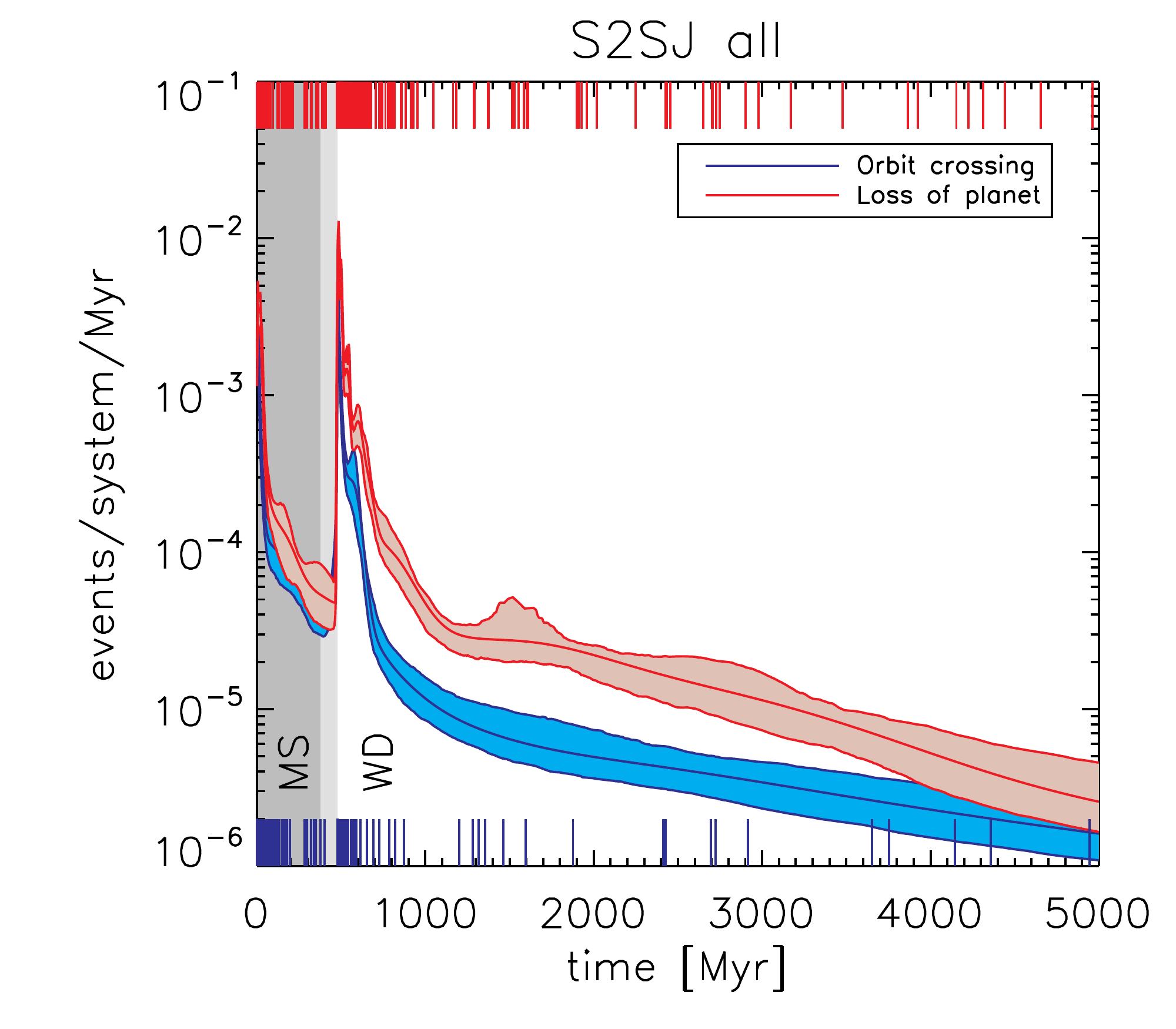}
   \caption{Kernel density estimates (and $1\sigma$ bootstrap 
     confidence intervals) showing times at which instability occurs 
     (here defined as the onset of orbit-crossing) 
     and planets are lost (to collision or ejection) 
     in preliminary integration sets (with no test particles). 
     \textbf{Top row: }All runs in sets with planet spacings from 
     $5-7$ mutual Hill radii. \textbf{Top left: }
     Super-Earth to Neptune-mass planets, set \textsc{se2n-57rh}. 
     \textbf{Top centre: }Neptune- to Saturn-mass planets, set \textsc{n2s-57rh}. 
     \textbf{Top right: }Saturn- to super-Jupiter mass planets, set \textsc{s2sj}.
     \textbf{Bottom row: }All sets (all separations) for Super-Earth to Neptune-mass planets 
     \textbf{(Bottom left)} and Saturn- to super-Jupiter mass planets 
     \textbf{(Bottom right).}
     Kernel density estimates are constructed with an adaptive-width ($k$-nearest 
     neighbour) Gaussian kernel. Individual times of events are shown as ticks on the upper 
     and lower axes. Background shading shows the star's evolutionary state, 
     as in Figure~\ref{fig:aqQ}. In all panels, instabilities occur at a decreasing rate 
     along the MS. Instabilities then spike following mass loss, after 
     which the rate decays again. With higher mass planets (\textsc{s2sj}), 
     the rate at which planets are lost closely tracks the rate at 
     which orbit-crossing commences, but lower-mass planets show 
     a more constant rate of planets being lost as it can take 
     many Gyr to change the orbits sufficiently to eject a planet 
     or force it to collide with the star.}
  \label{fig:notp}
\end{figure*}

 We now consider the distribution of stellar ages at which 
 instability occurs. Figure~\ref{fig:notp} shows this distribution 
 in different simulation sets. Each panel shows two 
 kernel density estimates\footnote{In this paper we show 
 kernel density estimates constructed with an adaptive-width 
 ($k$-nearest neighbour) Gaussian kernel 
 \citep{FeigelsonBabu12}. The adaptive width 
 resolves abrupt features, such as the sudden increase in 
 events after mass loss, while smoothing over regions 
 of lower density such as the tail at later times.}: one for the distribution 
 of times at which orbit-crossing begins, and one 
 for the distribution of 
 times at which planets are lost, be 
 that due to ejection or collision with each other or with the star. 
 The upper panels show the simulations with planets in the three mass 
 ranges and initially  spaced at $5-7$ mutual Hill radii, 
 some of the tightest spacings we considered. 
 These simulation sets show spikes in the rate of 
 the onset of orbit-crossing at very early 
 times on the main sequence and again following AGB mass loss. In the 
 \textsc{s2sj} set of massive planets, the times at which planets 
 are lost closely track the times of onset of orbit-crossing: 
 in these systems the instabilities quickly result in the 
 loss of one or more planets, usually to ejection. In contrast, 
 in the lower-mass simulation sets, the times at which planets are 
 lost follow a much flatter distribution, with no sudden spike 
 following AGB mass loss for the lowest mass planets (\textsc{se2n}). 
 Instabilities in these systems can take many Gyr to resolve, and many 
 unstable systems in which the orbits have begun crossing 
 actually retain all three planets for the whole integration. 
 The lower panels show the same distributions, but summed over all 
 initial separations for the \textsc{s2sj} and \textsc{se2n} mass 
 ranges. The timescale to begin orbit-crossing is a strong function 
 of separation, and so the incorporation of more widely-spaced 
 systems broadens the tail of systems which experience late instabilities. 
 Most instabilities, however, still begin at young WD cooling ages 
 of a few 100\,Myr. 
 The distribution of times at which planets are lost in the low-mass 
 \textsc{se2n} systems remains rather flat; we shall see that 
 this enables delivery of material to the WD for several Gyr 
 in these systems.

The orbital expansion of the planets during the AGB closely followed 
the expectation for the adiabatic regime that eccentricity should 
not be excited. We did however find a moderate eccentricity excitation 
of a few per mil. Single planets placed at $10$ and $20$\,au from our 
stars had their eccentricities excited to $0.0014$ and $0.0034$ 
respectively. Planets at larger separations will experience more 
eccentricity excitation. This can, for example, repopulate 
unstable chaotic regions surrounding a planet's orbit 
\citep{CaiazzoHeyl17}. \cite{PuWu15} found that, in terms of 
stability lifetime, increasing planetary eccentricity by $\sim0.01$ 
is roughly equivalent to reducing the separation by one mutual 
Hill radius. Planets on wider orbits than we have considered---a 
few 10s of au---will therefore experience destabilisation 
slightly more often.

Having conducted the simulations without test particles 
described in this Section, we select three 
individual runs for further study. We take one from each of 
\textsc{s2sj-57rh, n2s-57rh} and \textsc{se2n-57rh} and add in test particles, as 
described in the following Section. In each of these simulations the 
planets were stable throughout the main sequence but unstable after AGB mass loss, 
providing an ideal scenario for WD pollution. Parameters for our selected runs 
are given in Table~\ref{tab:selected}.

\begin{table}
  \begin{tabular}{lccc}
    Run & Planet mass $[\mathrm{M}_\oplus]$ & Semimajor axis [au] & Separation $r_\mathrm{H,mut}$\\
    \hline
    \textsc{s2sj} &  276.9 & 10.00 &  -  \\
                  &  462.3 & 14.58 & 5.95\\
                  &  121.8 & 20.14 & 5.52\\
    \\
    \textsc{n2s}  &   98.3 & 10.00 &  -  \\
                  &   65.2 & 12.63 & 6.12\\
                  &   19.5 & 15.39 & 6.46\\
    \\
    \textsc{se2n} &    1.3 & 10.00 &  -  \\
                  &   30.6 & 11.60 & 6.74\\
                  &    7.8 & 13.07 & 5.11
  \end{tabular}
  \caption{Planetary parameters (masses, radii, and separation in 
    mutual Hill radii) for the runs used in 
    Section~\ref{sec:tp} for the full integrations 
    including test particles.}
  \label{tab:selected}
\end{table}

\section{Main integrations: test particles included}

\label{sec:tp}

We now proceed to study the efficiency and rate of delivery 
of planetesimals to the WD by adding massless test particles 
to our simulations. The use of massless particles allows us 
to scale the simulation results to any belt mass, but requires 
that there be no significant gravitational effect of the 
planetesimals on the planets; most significantly, that 
there be no eccentricity damping. We show in the Appendix 
that this is satisfied if each planet is $\gtrsim10$ times 
more massive than the total planetesimal mass. For our lowest 
mass planets, this imposes a maximum belt mass of 
$\sim0.1\mathrm{\,M}_\oplus$.

The test particles in our simulations feel the gravitational 
force from the star and planets but do not experience 
non-gravitational forces. These can include gas drag 
from the stellar wind and 
stellar radiation forces such as Poynting--Robertson 
drag and the Yarkovsky effect, 
which can all be significant when the star approaches the 
AGB tip \citep{BonsorWyatt10,Dong+10,Veras16}. These forces 
are size-dependent and their inclusion would require 
specifying a size distribution for the planetesimals, 
in addition to the added computational cost. 
We plan to investigate the effects of non-gravitational forces in 
future work.

We take each of our chosen integrations and clone it 20 times, retaining 
the same input positions and velocities for the planets. We then 
add 200 test particles to each integration: 10 integrations 
with all of the test particles interior 
to the innermost planet (runs \textsc{*-inner}), 
and 10 with all of them exterior to the outermost planet (runs \textsc{*-outer}). 
We also run one simulation in which the test particles are distributed 
between the orbits of the innermost and the outermost planet 
(run \textsc{se2n-straddle}). 
In total we thus have 61 simulations and 12\,200 test particles. Simulations 
are again run for 5\,Gyr. Note that, as we use the adaptive-timestep RADAU algorithm, 
the evolution of the massive planets is not the same in each integration. In 
Mustill et al.\ (in prep) we describe the tests we conducted to 
verify that use of the RADAU integrator provides statistically robust results, 
even though each system's evolution is chaotic and affected by the integrator 
timestep (set by the orbits of the test particles and the close encounter history).

\begin{figure*}
  \includegraphics[width=\textwidth]{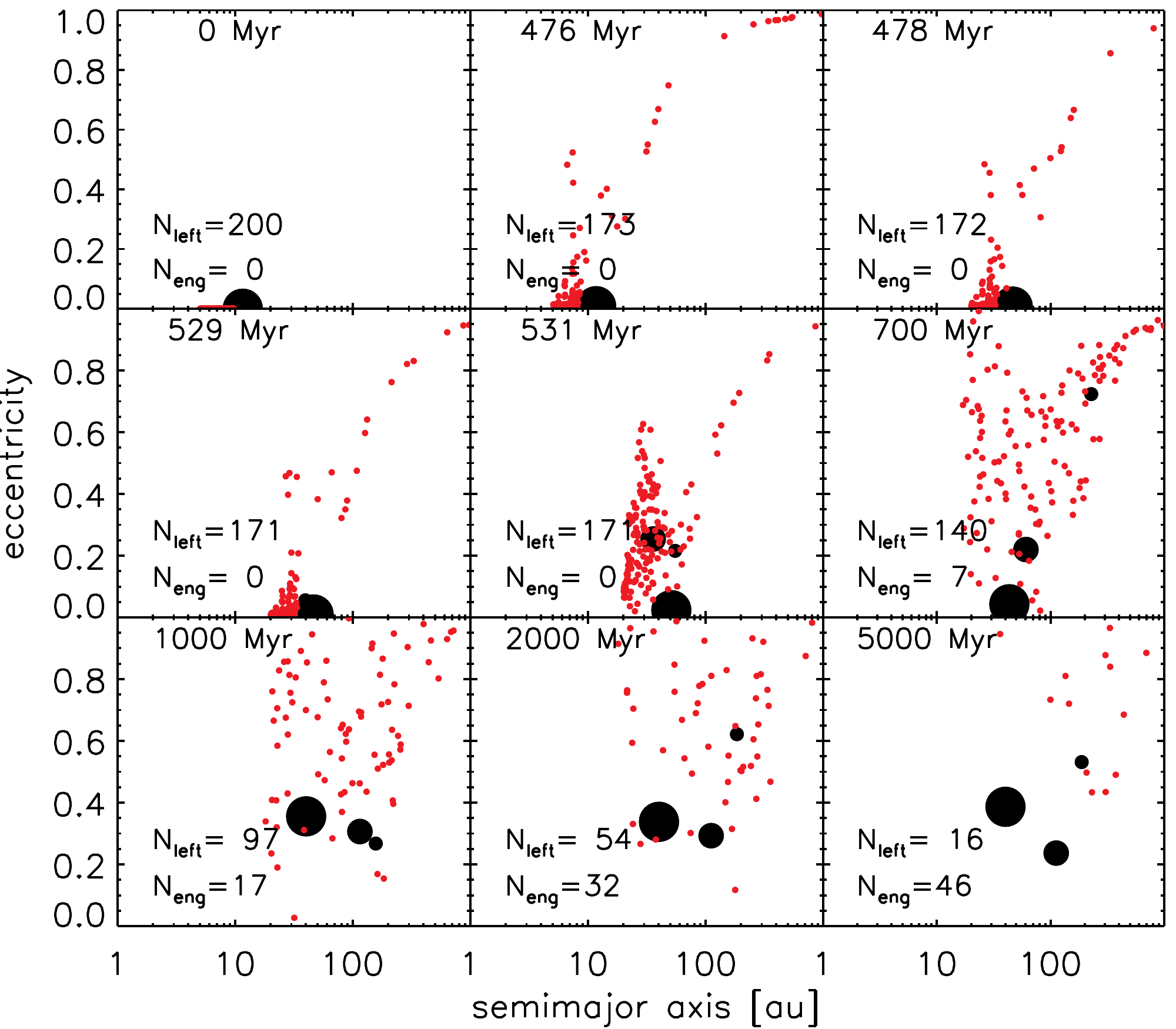}
  \caption{Time evolution of the run \textsc{se2n-inner-07} 
    with three low-mass planets (black, radius proportional 
      to cube root of mass) plus an inner belt of 200 test 
    particles (red). Each panel shows a different time snapshot; the 
    cumulative number of particles to have been engulfed into the WD's 
    atmosphere ($N_\mathrm{eng}$) and the number of particles 
      remaining ($N_\mathrm{left}$) are shown at the bottom of each panel. 
    \textbf{Top left: }Initial state. 
    \textbf{Top centre: }State just before significant mass loss at the AGB tip. 
    Some belt particles have been scattered or ejected. 
    \textbf{Top right: }State after AGB mass loss. The entire system has expanded 
    almost adiabatically.
    \textbf{Middle and bottom rows: }Development of the planetary 
    instability and subsequent clearing of the belt. Orbit-crossing begins at 
    around $530$\,Myr but all planets remain bound until the end of the integration 
    at 5\,Gyr. During this time 46 test particles (out of 173 surviving 
    to the WD phase) collide with the WD, as their eccentricity is 
    excited to near unity while their semi-major axes remain at a few 
    tens of au.}
  \label{fig:snapshots}
\end{figure*}

We set the limits of the planetesimal belts to slightly intrude 
into the chaotic zone of the innermost or outermost planet 
\citep{Wisdom80}. 
This allows for a natural sculpting of the belt over the 
star's main sequence lifetime, as well as for 
the natural expansion of the chaotic zone following AGB 
mass loss but before planetary instability 
\citep{Bonsor+11,FrewenHansen14}. Particles are distributed 
uniformly in semi-major axis within these belts. Limits 
on the initial belt semimajor axes are given in Table 
\ref{tab:nengulf}. The two righthand panels of Figure~\ref{fig:setup} 
illustrate the setup, and an example initial configuration in 
$a-e$ space can be seen in the top left panel of Figure~\ref{fig:snapshots}.

\subsection{Time evolution of planetesimal belts}

\begin{figure*}
  \includegraphics[width=0.33\textwidth]{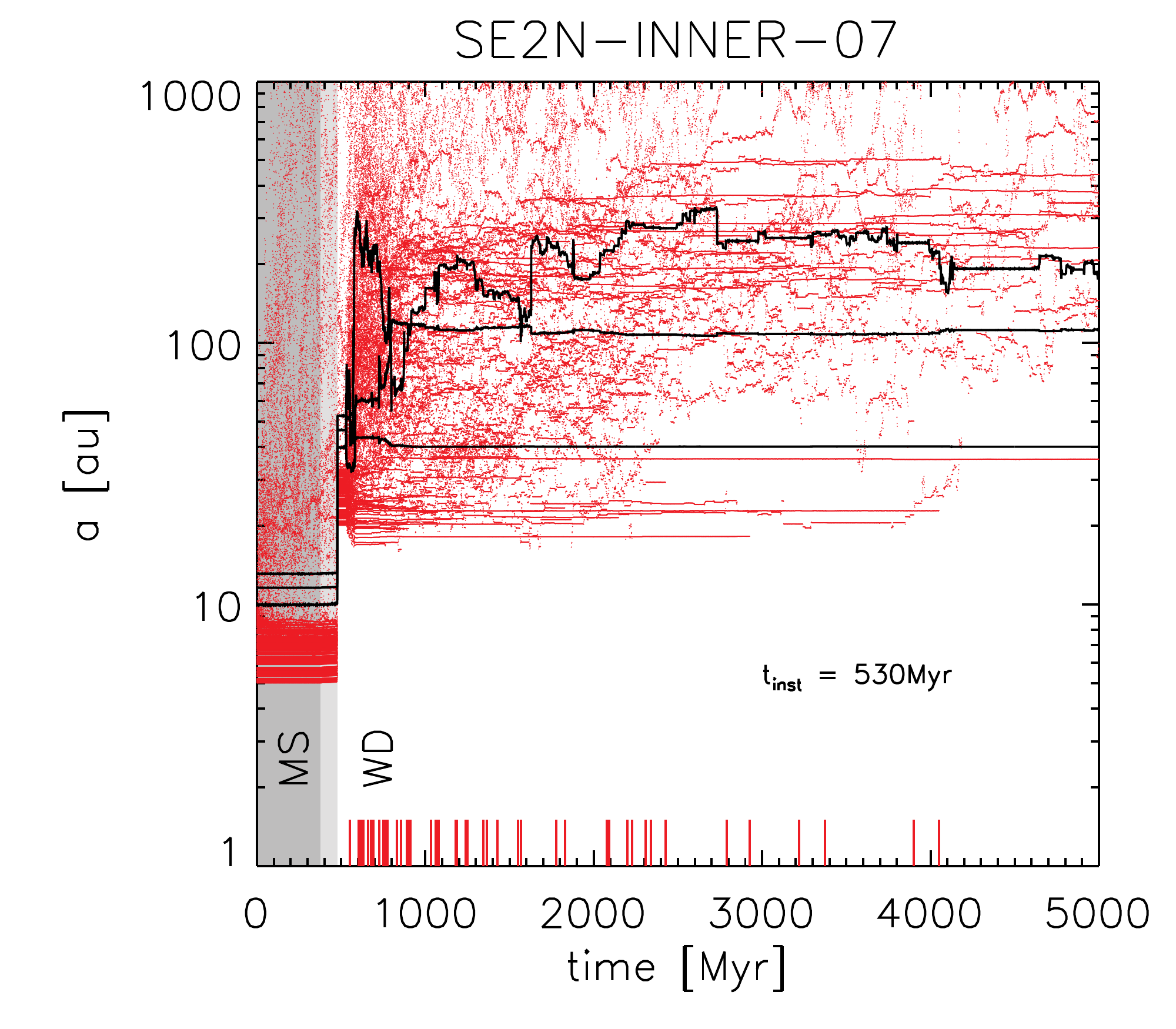}
  \includegraphics[width=0.33\textwidth]{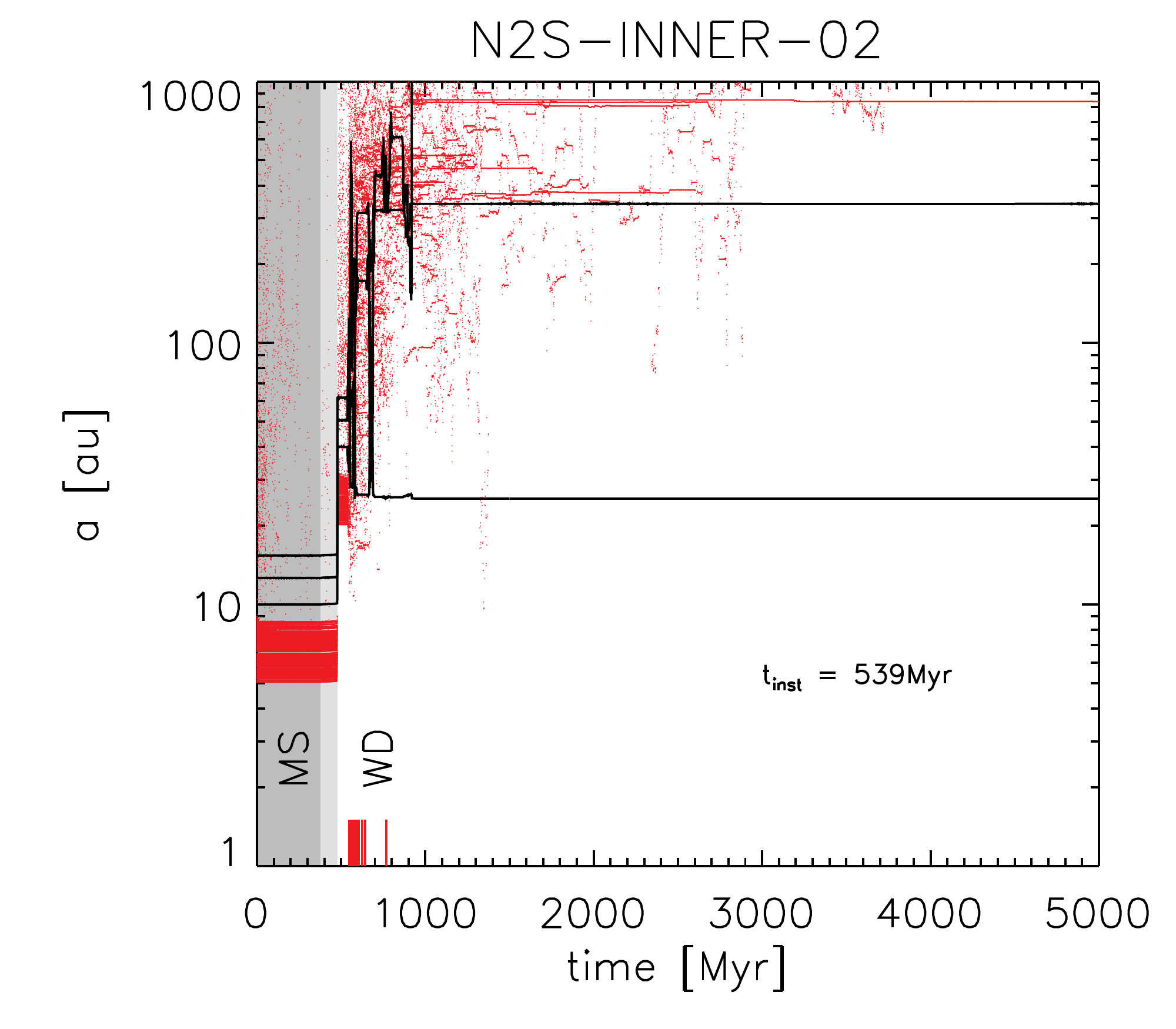}
  \includegraphics[width=0.33\textwidth]{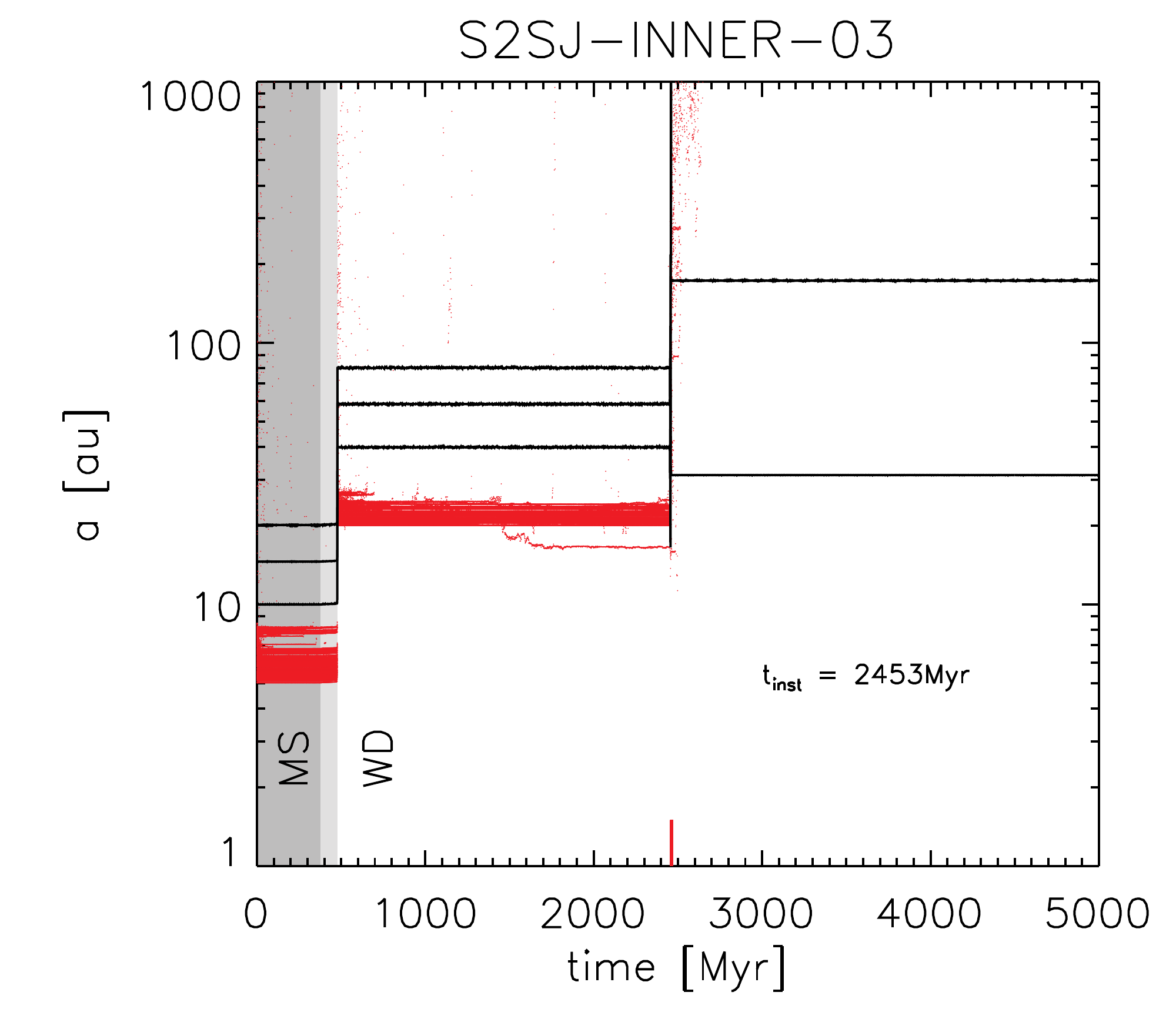}
  \includegraphics[width=0.33\textwidth]{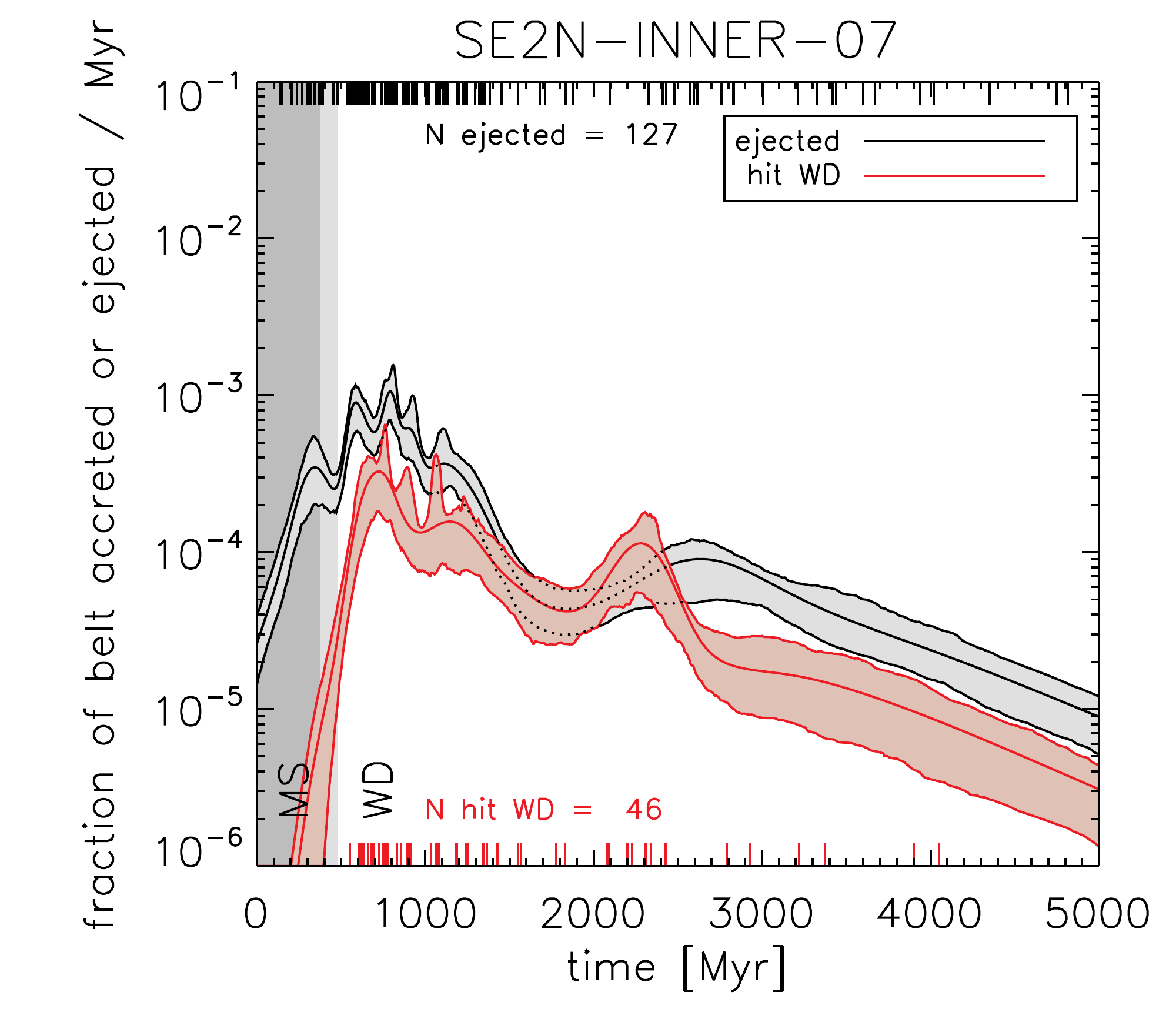}
  \includegraphics[width=0.33\textwidth]{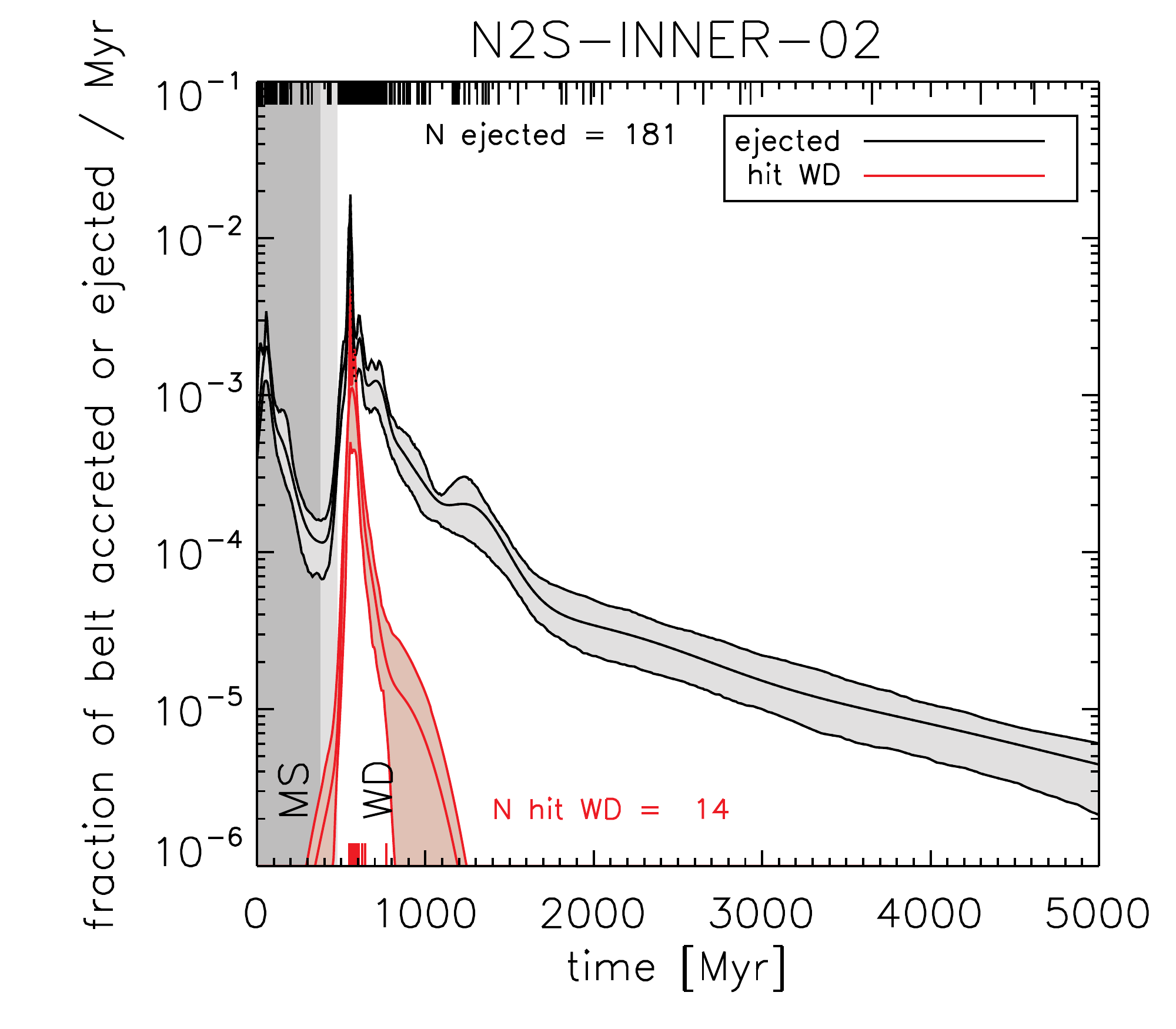}
  \includegraphics[width=0.33\textwidth]{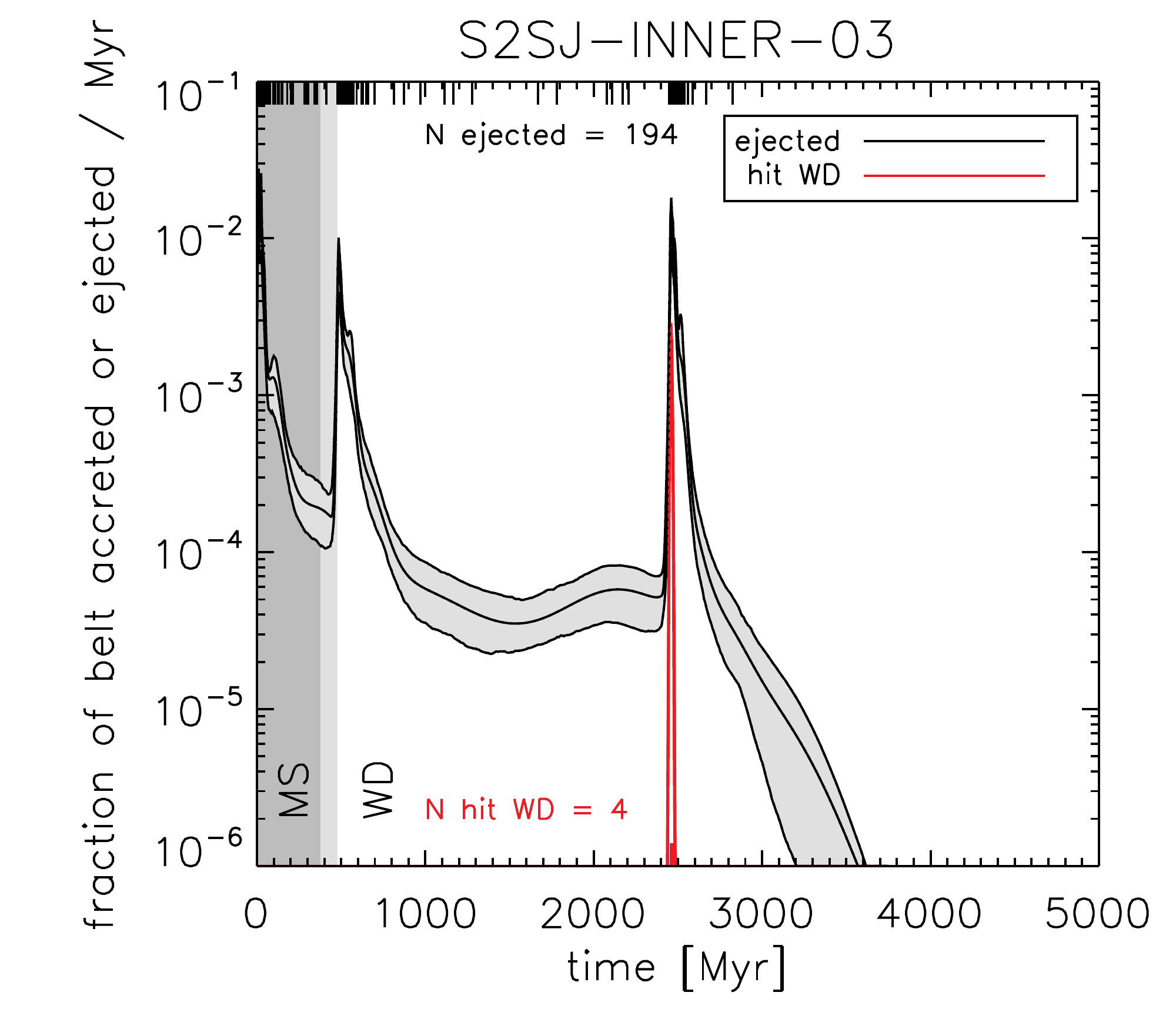}
  \caption{Time evolution of some example systems. \textbf{Top: }Planets' 
    semimajor axes are marked in black, planetesimals' in red. Large 
    ticks at the bottom axis mark where test particles collided 
    with the WD. The time at which orbit-crossing begins, 
    measured from the zero-age main sequence, is shown on the plot. 
    \textbf{Left: }\textsc{se2n} example. This is the run that resulted 
    in the greatest number of planetesimals hitting the WD (46 in all). 
    Scattering amongst the planets begins around 50\,Myr after 
    the star becomes a WD (total age 530\,Myr), but all three planets remain 
    bound for the duration of the simulation. This allows material to 
    be delivered to the WD for several Gyr, albeit at a decreasing rate. 
    \textbf{Centre: }\textsc{n2s} example. An instability among the 
    planets soon after mass loss results in delivery of material to 
    the WD over a period of a few hundred Myr. Following ejection 
    of one of the planets, remnant belt particles are slowly cleared, 
    but none at these late times collide with the WD. 
    \textbf{Right: }\textsc{s2sj} example. Here we see a widening 
    of the chaotic zone and ``Kirkwood gaps'' along the main sequence, 
    a further broadening of the chaotic zone following mass 
    loss, and then an instability among the planets just before 
    $2.5$\,Gyr. This rapidly depletes the rest of the belt, and causes 
    a small amount of material to be delivered to the WD. The final 
    system comprises two planets and no test particles. \textbf{Bottom: }
    Kernel density estimates 
    of times at which planetesimals hit the star or are ejected, with 
    times marked as large tick marks on the upper and lower axes. 
    The rate at which particles are lost increases dramatically 
    following stellar mass loss and the onset of orbit-crossing 
    amongst the planets.}
  \label{fig:evolution}
\end{figure*}

Figure~\ref{fig:snapshots} shows snapshots of one system at 
different stages in its evolution. The system begins with three 
low-mass planets and an interior belt of 200 test particles, 
all bodies being on circular orbits. At 476\,Myr, 
as the star approaches the 
AGB tip, the outer region of the belt has been dynamically 
eroded. Mass loss over the next 2\,Myr causes the 
entire system to expand outwards. The mass loss 
also destabilises the planets, causing orbit-crossing to 
begin at around 530\,Myr. The planets continue to scatter 
while remaining bound for the next $4.5$\,Gyr, during which 
time the belt is slowly depleted with 46 particles (out of 
173 surviving MS and AGB evolution) colliding with the 
WD---most particles are ejected from the system. 
After 5\,Gyr the three planets remain in the 
system, along with 16 test particles.

An alternative view of the same system, together 
with examples from the other two planetary mass ranges, 
is shown in Figure~\ref{fig:evolution}. Here we show 
the semi-major axes of all bodies in the system 
(upper panels), and the distribution of times at 
which particles were ejected or hit the star 
(lower panels). 
The belts typically go through four phases of evolution:
\begin{enumerate}
\item Pre-WD evolution: unstable regions of the belt 
  (the chaotic zone and unstable resonances) are cleared 
  as the system proceeds along the MS and up the giant 
  branches. These destabilised particles are typically 
  ejected, although with low-mass planets they may not have 
  time to escape the system as the clearing proceeds more 
  slowly \citep{MorrisonMalhotra15}.
\item WD-stage evolution before planetary instability: chaotic, unstable 
  regions of the belt expand following mass loss. This phase 
  is brief in the first two panels of Figure~\ref{fig:evolution}, 
  but is clearly seen in the right-hand panel. As before, 
  the destabilised particles are typically ejected. In 
  fact, no belt particles collided with the WD 
  between the end of the AGB and the onset of planetary orbit-crossing.
\item WD-stage evolution during planetary instability: 
  once planetary scattering begins, the original structure 
  of the belt is swiftly destroyed. This phase is of brief 
  duration when the planets are of high mass, but some 
  of the lower-mass systems (such as that shown in 
  the right-hand panel of Figure~\ref{fig:evolution}) 
  remain in this phase for the remainder of the integration. 
  This phase typically sees the highest rates of 
  delivery of material to the WD.
\item WD-stage evolution after planetary instability: 
  if the planets settle down into a stable configuration 
  following scattering, it may still take many Gyr 
  to deplete the belt. An example is seen in the middle panel of 
  Figure~\ref{fig:evolution}; in this example, the 
  particles destabilised in this phase were all ejected. 
  However, in some of the \textsc{se2n} integrations with lower-mass planets 
  delivery of material to the WD continues during this phase.
\end{enumerate}

These four phases are particularly evident in the right-hand 
panels of Figure~\ref{fig:evolution}. Rates of ejection begin 
high, and fall as the unstable regions of the belt are cleared along the 
main sequence. Ejections spike and fall again after the star 
becomes a WD. At around $2.4$\,Gyr, the planets undergo 
a short-lived instability resulting in the ejection of one 
of them. During and after this, ejection rates again peak and decay, 
and a handful of particles collide with the WD.

\subsection{Particles colliding with the WD}

\begin{table*}
  \begin{center}
    \begin{tabular}{lcccccccccc}
      Simulation set           & $N_\mathrm{runs}$ & $N_\mathrm{particles}$ & $a_\mathrm{in}$\,[au] & $a_\mathrm{out}$\,[au] & $N_\mathrm{surv, WD}$ & $N_\mathrm{surv, 5Gyr}$ & $N_\mathrm{engulf}$ & $f_\mathrm{engulf}$ & $N_\mathrm{Roche}$ & $f_\mathrm{Roche}$\\
      \hline
      \textsc{S2SJ-outer}      &  8$^1$& 1600  & 25.0 & 45.0 & 1517  &    27  &  11  &  0.7\% &   72 &  4.7\%\\
      \textsc{S2SJ-inner}      & 10    & 2000  &  5.0 &  8.5 & 1069  &     6  &  25  &  2.3\% &  142 & 13.3\%\\
      \textsc{N2S-outer}       & 10    & 2000  & 16.5 & 25.0 & 1576  &    95  &  85  &  5.4\% &  342 & 21.7\%\\
      \textsc{N2S-inner}       & 10    & 2000  &  5.0 &  9.0 & 1568  &    64  & 190  & 12.1\% &  585 & 37.3\%\\
      \textsc{SE2N-outer}      & 10    & 2000  & 13.5 & 25.0 & 1785  &   184  & 138  &  7.7\% &  362 & 20.3\%\\
      \textsc{SE2N-inner}      & 10    & 2000  &  5.0 &  9.8 & 1670  &   195  & 286  & 17.1\% &  620 & 37.1\%\\
      \textsc{SE2N-straddle}   &  1    &  200  & 10.0 & 13.1 &   58  &     9  &   2  &  3.4\% &    8 & 13.8\%
    \end{tabular}
    \caption{Fates of the test particles in our simulation sets. We give the
      number of runs for each configuration ($N_\mathrm{runs}$); the initial number
      of particles summed across those runs ($N_\mathrm{particles}$); the inner and
      outer edges of the planetesimal belt ($a_\mathrm{in}$ and $a_\mathrm{out}$); 
      the number of particles
      surviving when the star becomes a WD ($N_\mathrm{surv}$); the number
      colliding with the WD ($N_\mathrm{engulf}$) and the
      fraction of the surviving particles that these represent ($f_\mathrm{engulf}$); 
      and the number entering the Roche limit of the WD ($N_\mathrm{Roche}$) and the 
      fraction of surviving particles that these represent ($f_\mathrm{Roche}$).
      $^1$In two of the ten \textsc{S2SJ-wide-outer} runs the planets were stable and
      we ignore them when compiling these statistics. No particles struck the 
        WD in these runs.}
    \label{tab:nengulf}
  \end{center}
\end{table*}

\begin{figure*}
  \includegraphics[width=0.33\textwidth]{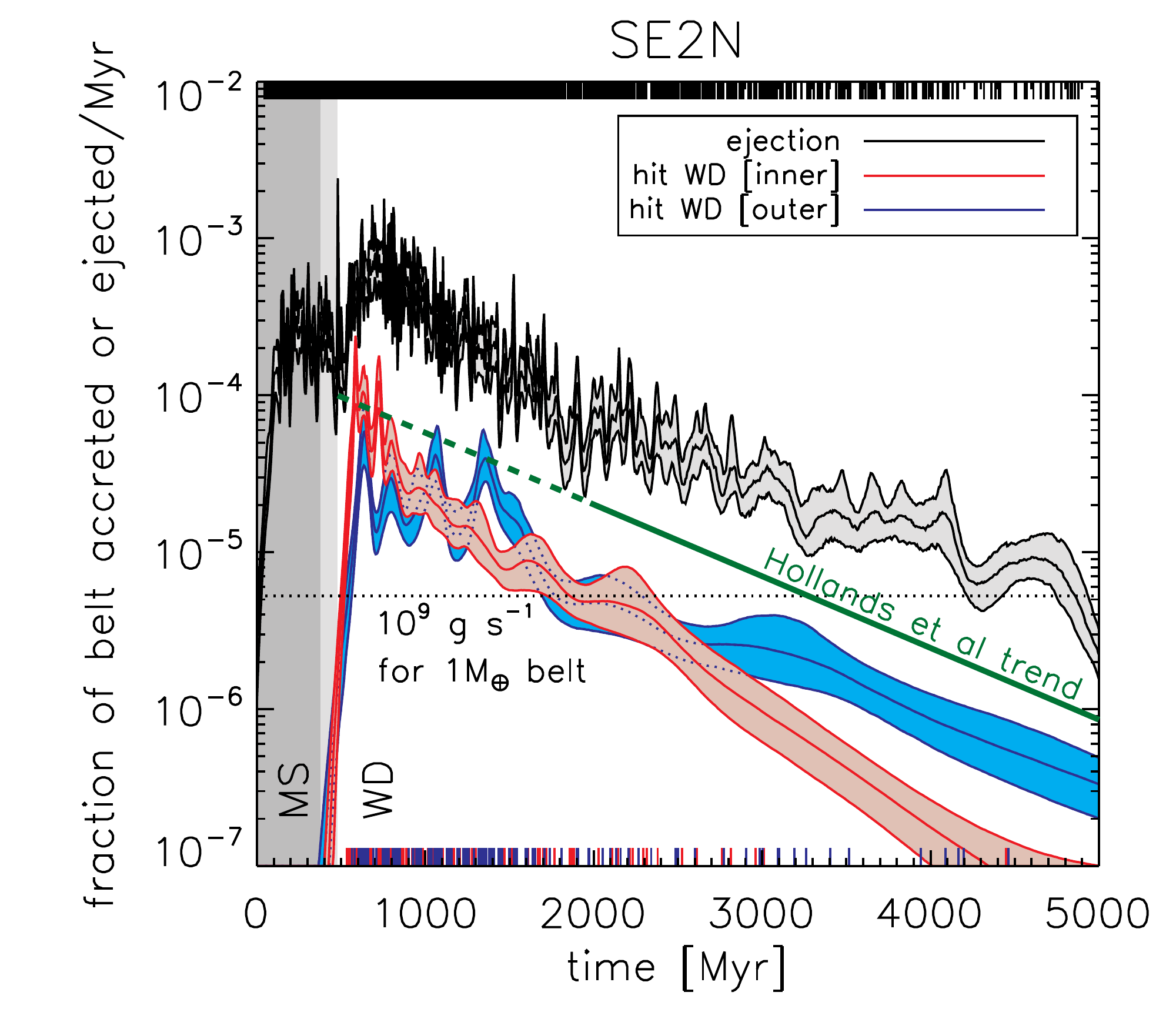}
  \includegraphics[width=0.33\textwidth]{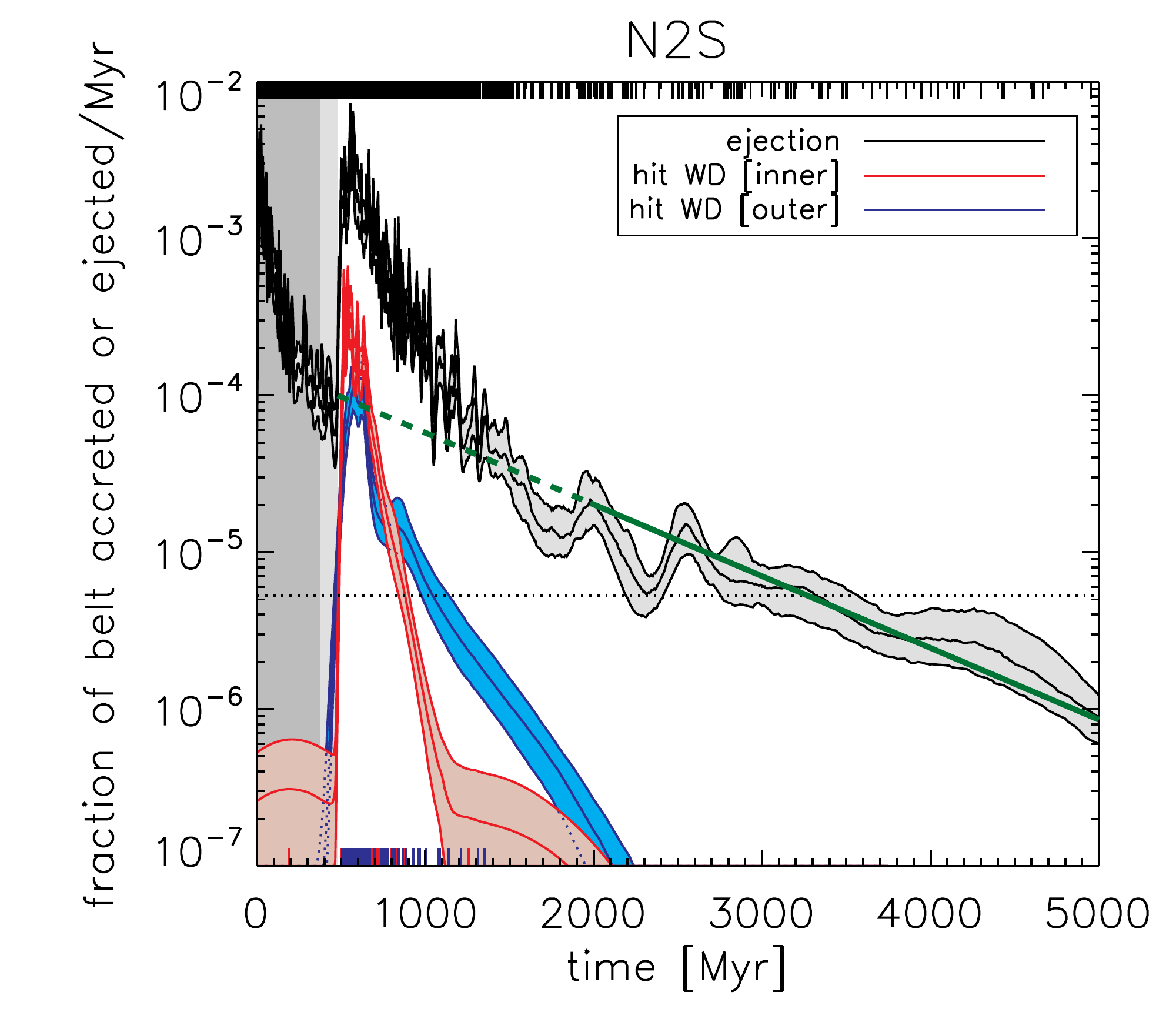}
  \includegraphics[width=0.33\textwidth]{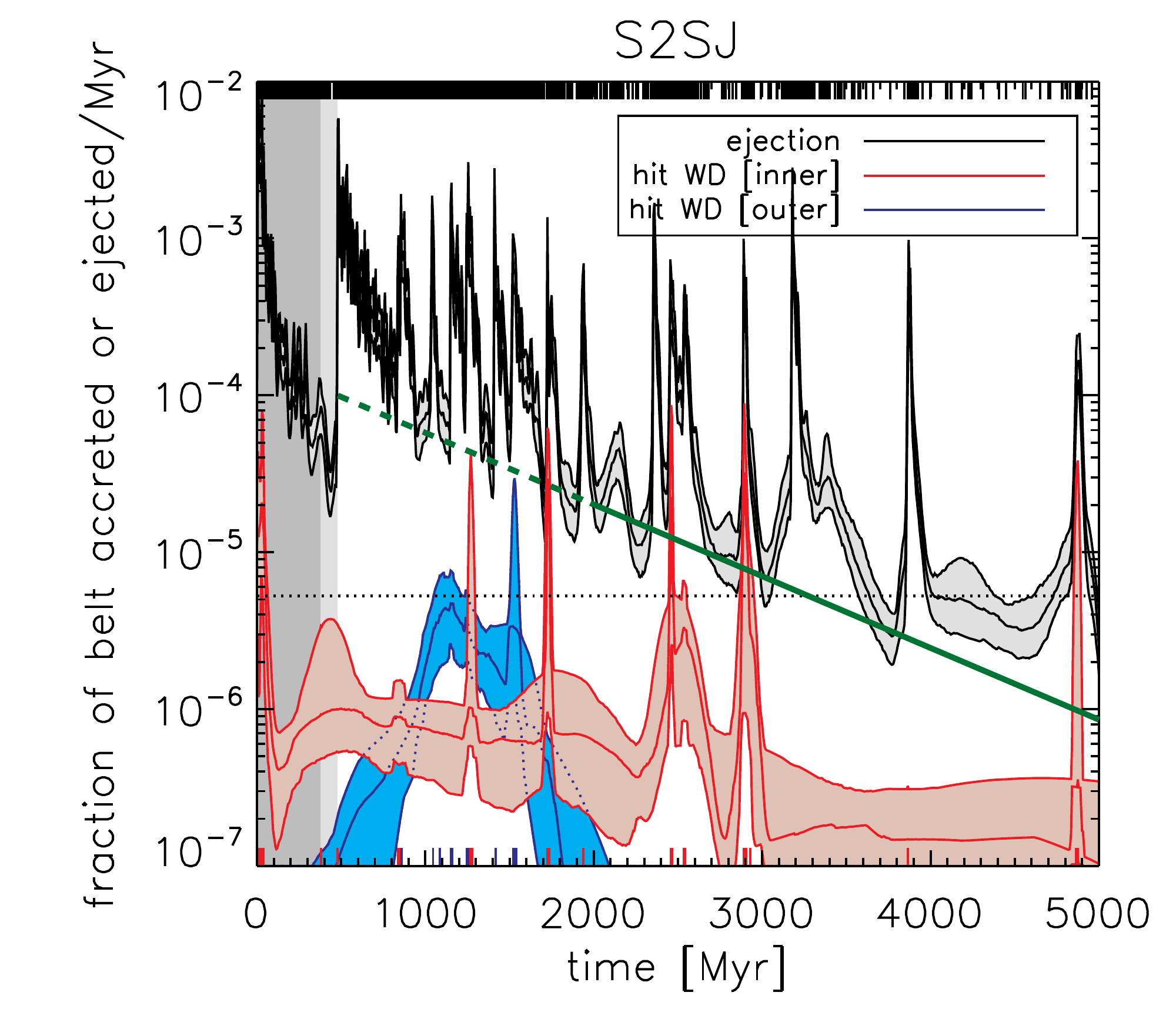}
  \includegraphics[width=0.33\textwidth]{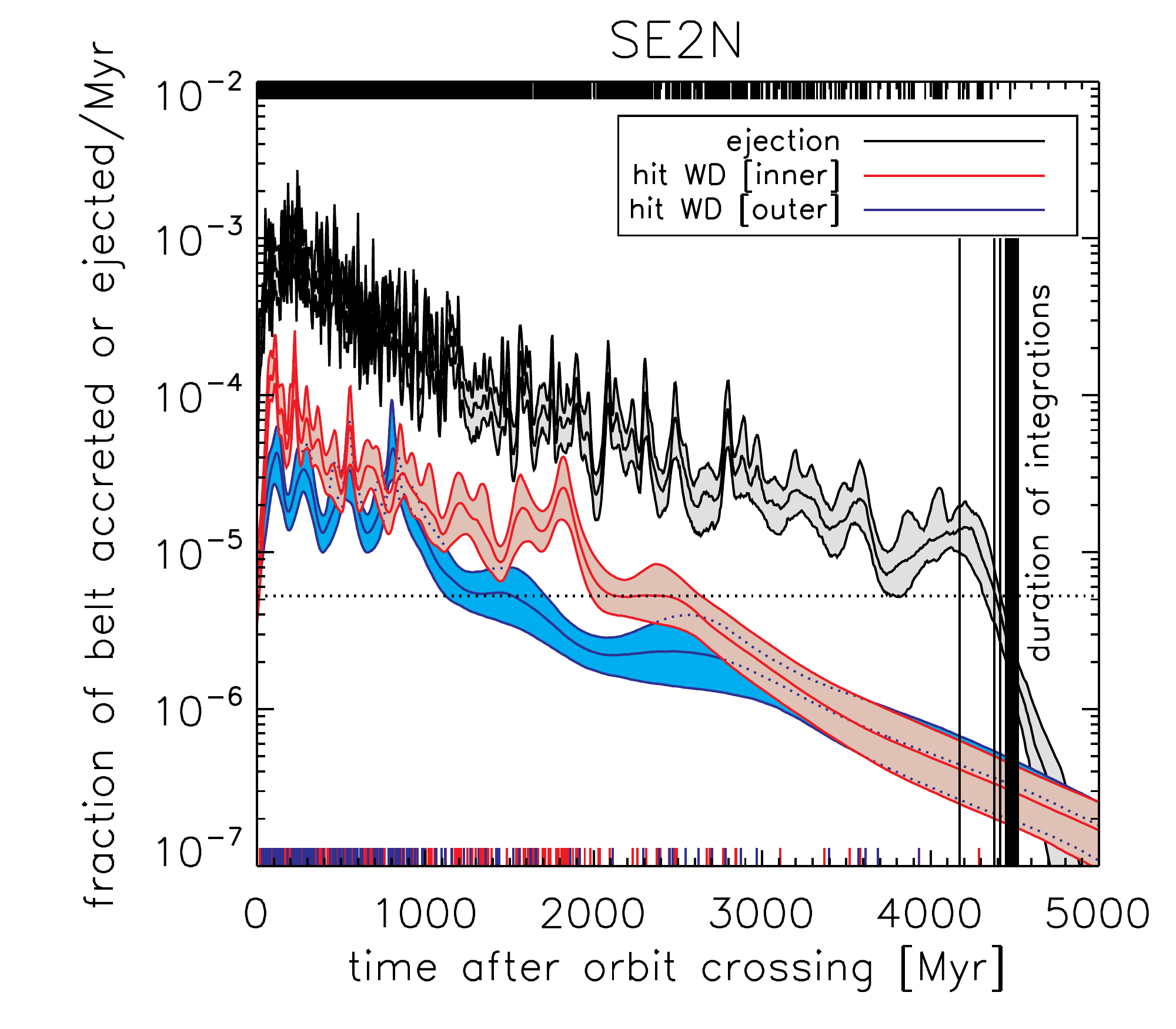}
  \includegraphics[width=0.33\textwidth]{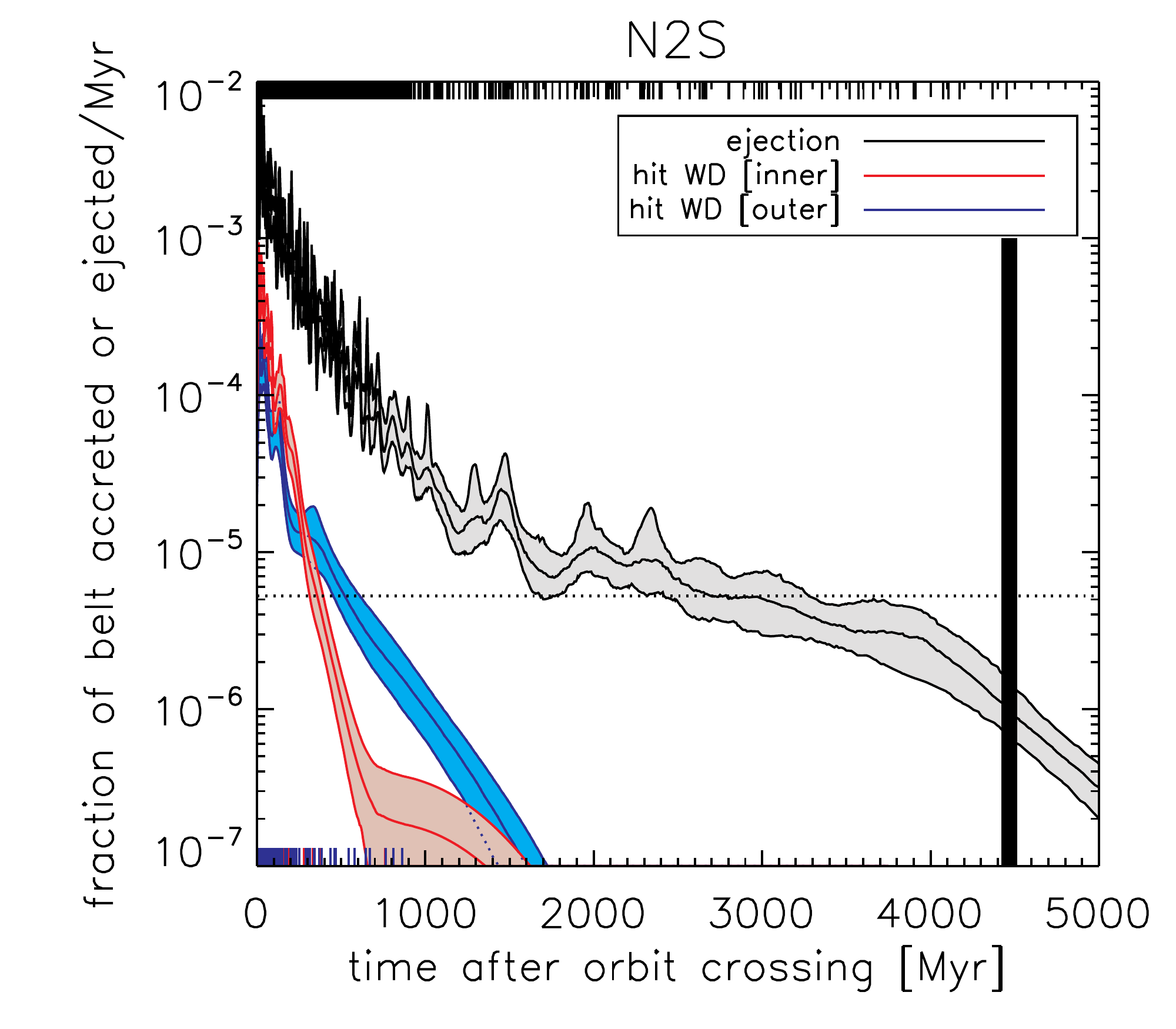}
  \includegraphics[width=0.33\textwidth]{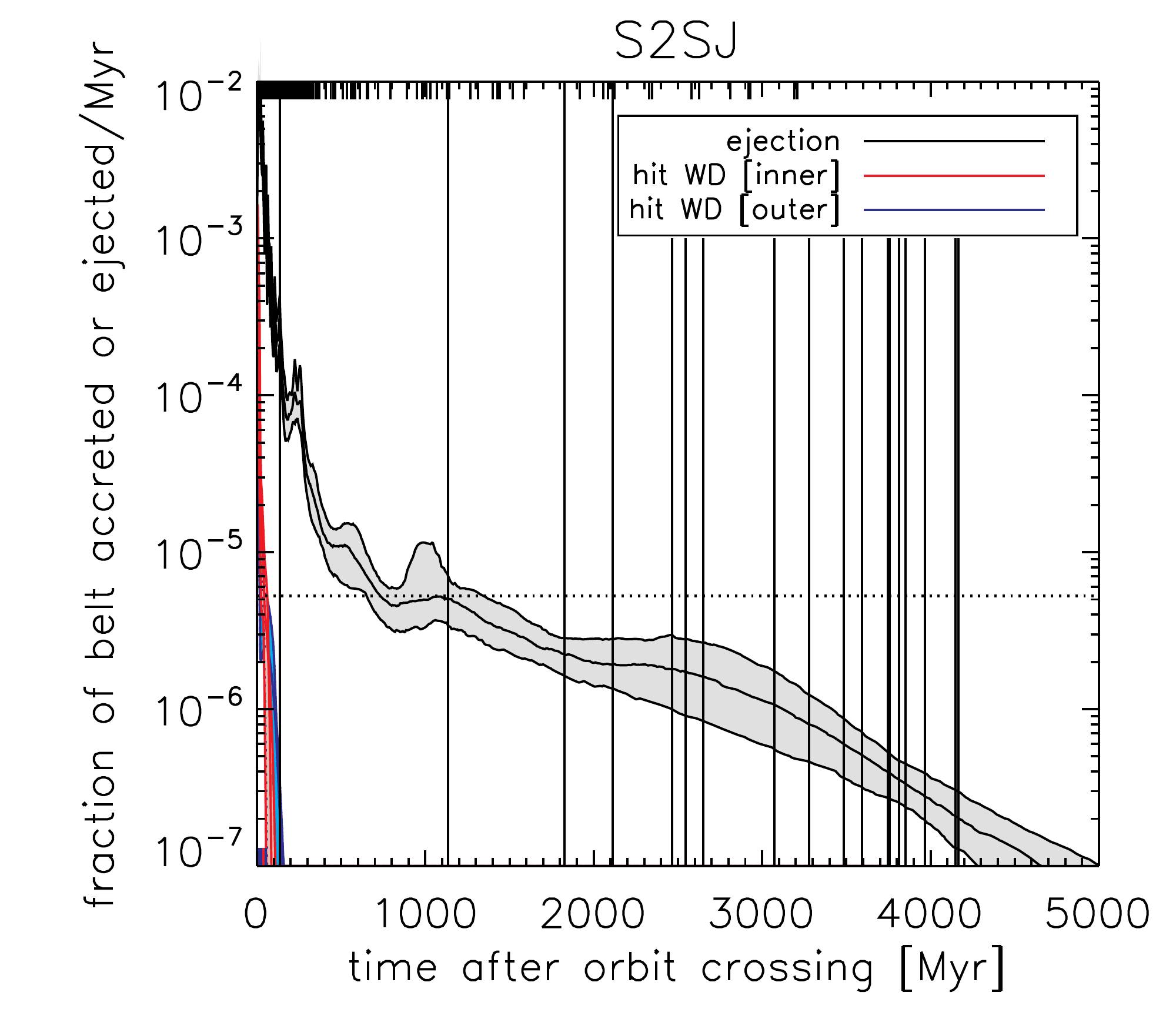}
  \caption{\textbf{Top row: }Kernel density estimates of 
    the times at which test particles 
    hit the star or are ejected from the system. 
    Ejections are summed across all integrations in a given 
    set; stellar collisions are divided according to whether the particles 
    originate in an inner or an outer belt. 
    The actual times of ejection are marked on the upper horizontal 
    axis as black ticks; times of stellar collision are marked on the lower 
    horizontal axis with red ticks (if from the inner belt) 
    or purple ticks (if from the outer belt). 
    We also show as a horizontal dotted line an accretion 
    rate of $10^9\mathrm{g\,s}^{-1}$ for a $1\mathrm{\,M}_\oplus$ 
    belt. The solid green line shows the observational 
    trend (with arbitrary normalisation) from Hollands 
    et al (submitted) for late cooling 
    ages of over $1.5$\,Gyr, together with its extrapolation to 
    younger cooling ages. 
    \textbf{Bottom row: }
    As above, but showing the time of ejection/collision after 
    planetary instability (orbit-crossing) begins: 
    each system is separately translated so that its 
    orbit crossing begins at $t=0$, and then the 
    kernel density estimate is constructed. 
    Vertical lines mark the durations of individual 
    simulations past the onset of orbit crossing.
  }
\label{fig:tengulf}
\end{figure*}

We now turn our attention to the particles colliding with the WD, and the 
times at which this occurs. The 
numbers of particles directly hitting the WD in each simulation set, 
as well as the number crossing the Roche limit, 
are shown in Table~\ref{tab:nengulf}. 
This table also shows the number of the particles remaining in 
the integration when the star became a WD, the number that 
survive to the end of the integration, and the fraction of the
belt surviving when the WD formed that later collided with the WD. 
Two trends are clear from this table: particles in the inner 
belts are more likely to hit the star than particles in outer 
belts, and the lower-mass planets are much more efficient 
at causing accretion as opposed to ejection of particles. 
Ejection and collision with the star were the dominant loss 
channels after the star became a WD: only 21 particles across all 
simulations struck one of the planets after the star became a WD.

Here we consider the statistics of particles directly hitting 
the WD; as we shall discuss in Section~\ref{sec:collide}, particles cross the 
Roche limit at a higher rate but with the same qualitative trends. 
Kernel density estimates of the times 
at which particles are lost to stellar collision or ejection in three example integrations 
are shown in the lower panels of Figure~\ref{fig:evolution}. Both the 
number of particles hitting the star and the timescale over which this delivery  
is maintained increase as the mass of the planets decreases. 
To explore the rates of delivery as a function of time 
in a population of systems, all the integrations in each 
simulation set are then combined to give 
Figure~\ref{fig:tengulf}. Here, the upper panels show 
the absolute times at which particles were lost, while the lower panels 
show times relative to the onset of orbit-crossing among the planets. 
We also show for comparison the trend in the upper envelope of 
accretion rates from the DZ sample of Hollands et al (submitted), 
which decays exponentially with an e-folding time of $0.95$\,Gyr; 
additionally, we show an accretion rate of $10^9\mathrm{\,g\,s}^{-1}$ 
from a $1\mathrm{\,M}_\oplus$ belt, 
at the high end of WD accretion rates \citep{Koester+14}. 
We now discuss the simulation sets in more detail.

\subsubsection{Saturn to super-Jupiter}

These runs saw the lowest numbers of particles 
colliding with the WD, with only 11 from the outer 
belts and 25 from the inner belts being lost this way, representing 
respectively $0.6\%$ and $2.3\%$ of the particles 
surviving at the onset of the WD phase. 

Times at which particles strike the WD range from cooling ages of 300\,Myr to 4.4\,Gyr 
\footnote{This partly reflects the highly chaotic original system, whose stability lifetime 
is sensitive to the changes to the integrator timestep induced by 
the test particles.}. While this gives a respectable range, in any one simulation the accretion 
events usually occur in a relatively brief window of a few 10s of Myr following 
instability among the giant planets, as can be seen in the lower right panels of 
Figures~\ref{fig:evolution} and~\ref{fig:tengulf}. Hence, instability among giant planets 
is unlikely to allow ongoing accretion for 100s of Myr to Gyr throughout 
the WD lifetime.

\subsubsection{Neptune to Saturn}

The intermediate-mass systems were much more efficient at 
delivering material to the WD. 85 particles from the outer 
belts and 190 from the inner belts struck the WD, representing 
respectively $5.4\%$ and $12.1\%$ of the particles
surviving at the onset of the WD phase.

The time of the instability amongst the planets in these simulation was 
not so variable as in the \textsc{s2sj} set.
Compared to the more massive planets, run \textsc{n2s} shows a broader peak 
in the distribution of times of delivery. Sporadic ejections 
continue for several Gyr. However, 
the accretion is still restricted to a relatively brief window 
of a few 100 Myr during and after instability 
(centre panels of Figures~\ref{fig:evolution} and~\ref{fig:tengulf}).

\subsubsection{Super-Earth to Neptune}

These runs were the most efficient at delivering particles 
to the WD, with 138 particles from outer belts and 286 
from inner belts colliding with the WD, representing respectively 
$7.7\%$ and $17.1\%$ of the particles
surviving at the onset of the WD phase.

This run shows a still broader distibution of 
delivery times (left-hand panels of 
Figures~\ref{fig:evolution} and~\ref{fig:tengulf}). 
Delivery is maintained even at late times 
of several Gyr (measured both 
absolutely and with respect to the onset of planetary instability), 
although the rate does slowly decay. High rates of $\gtrsim10^{-5}$ 
of the belt per Myr are maintained for the first 2\,Gyr after 
instability. This rate represents $\sim2\times10^{9}\mathrm{\,g\,s}^{-1}$ 
for a belt of $1\mathrm{\,M}_\oplus$. The rate of decay in the 
upper envelope of observed accretion rates at late times 
found by Hollands et al (submitted) is matched very well by the 
simulations with outer belts, while the inner belts have an 
accretion rate that falls off a little more steeply. 
Both of the belt configurations agree much more closely 
with the observed trend than do any of the configurations 
with higher-mass planets.

\subsubsection{Collision with the WD \emph{versus} tidal disruption}

\label{sec:collide}

\begin{figure}
  \includegraphics[width=0.5\textwidth]{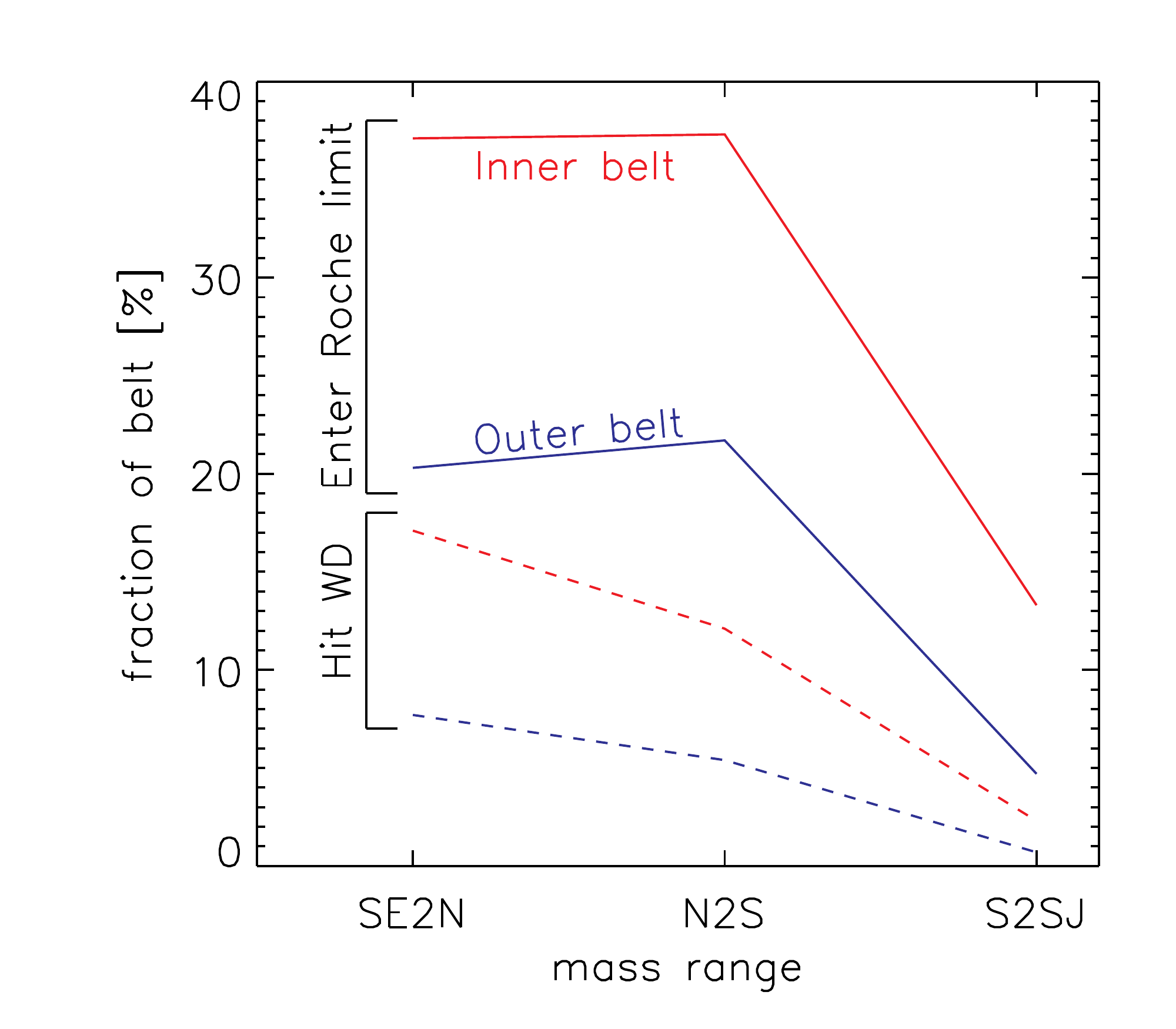}
  \includegraphics[width=0.5\textwidth]{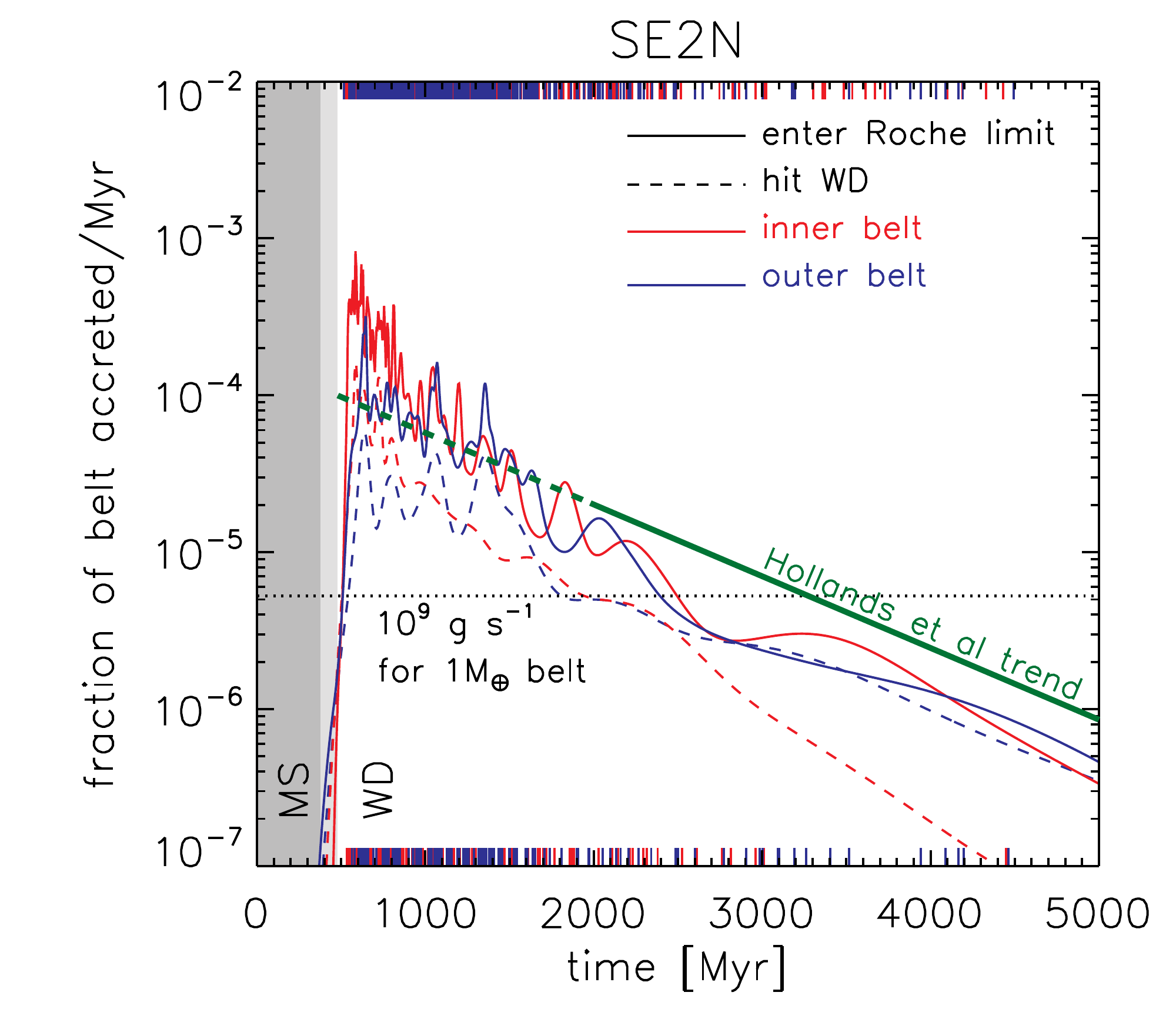}
  \caption{\textbf{Top:} Accretion efficiency onto the WD in our simulations. 
    The horizontal axis shows the mass range of the planets: 
  \textsc{se2n} ``super-Earth to Neptune''; \textsc{n2s} 
  ``Neptune to Saturn'', and \textsc{s2sj} ``Saturn to 
  super-Jupiter''. On the vertical axis is plotted 
  the fraction of the belt which survived AGB evolution 
  which later enters the WD's Roche limit ($0.003$\,au, 
  solid lines) or collided with the WD ($5\times10^{-5}$\,au, 
  dashed lines). The red and blue lines mark simulations with 
  inner and outer belts respectively. The efficiency of 
  delivery increases with decreasing planet mass. 
  \textbf{Bottom:} Rates of delivery to the Roche limit 
  (solid) and the WD radius (dashed) in our \textsc{se2n} 
  runs. Red and blue lines mark simulations with
  inner and outer belts respectively. Despite the higher 
  efficiency with which material is delivered to the Roche limit 
  compared to the WD surface, the trend with time is the same.}
  \label{fig:f_acc}
\end{figure}

Hitherto we have considered a conservative condition for 
the accretion of material onto the WD: the physical 
collision between a particle and the star. However, 
the physical radius of the WD is smaller, by a factor of 
$\sim100$, than the Roche limit at which bodies are 
tidally disrupted. Bodies with pericentres smaller than 
this distance (roughly 1 Solar radius, depending on density) 
will be shredded into long debris trails 
\citep[e.g.,][]{Debes+12,Veras+14}. An \emph{optimistic} 
condition for accretion onto the WD would be to assume that 
all material entering this Roche limit would be accreted. Thus, 
direct collision and entry inside the Roche limit should 
bracket the true amount of material accreted.

We show in the final two columns of Table~\ref{tab:nengulf} 
the total number of particles entering within $0.003$\,au 
of the WD, together with the fraction of the belt surviving 
the AGB that this represents. These values are considerably 
higher than those for the number of particles colliding 
with the WD: $4.7\%$ and $13.3\%$ for the \textsc{s2sj} sets 
with outer and inner belts respectively, compared to 
$0.7\%$ and $2.3\%$ for the particles actually hitting the WD. 
With the less massive planets, these values are $20\%$ and 
$37\%$ from the outer and inner belts. Interestingly, 
the \textsc{n2s} and \textsc{se2n} sets are equally 
efficient at driving material inside the Roche limit, 
in contrast to their efficiencies at delivering material 
to the WD radius where the \textsc{se2n} simulations 
perform better. The fraction of belt material delivered 
to the Roche limit and to the WD radius in the 
different simulation sets is displayed graphically in 
the upper panel of Figure~\ref{fig:f_acc}.

Material is delivered to the Roche radius at higher 
rates than it is delivered to the WD surface, as is 
to be expected since not all particles crossing 
the Roche limit will have their pericentres forced down 
to the WD radius itself. However, the trend in accretion rate 
with time, as measured by the rate at which material 
reaches these two thresholds, is broadly similar, 
as can be seen in the lower panel of Figure~\ref{fig:f_acc}
for the \textsc{se2n} simulations.  
At very late times, the accretion rate from the inner belts 
does not fall off as steeply when using the Roche limit as a 
threshold compared to when using the WD radius, and better 
matches the observed decay rate from Hollands et al (submitted). 
In the \textsc{n2s} and \textsc{s2sj} sets, the time 
over which material is delivered to the Roche limit 
is brief.

What is the fate of the material that crosses the Roche limit? 
In pure dynamical terms, some fraction of it 
does later collide with the WD: $17\%$ in the 
\textsc{s2sj} runs, $30\%$ in \textsc{n2s} and $43\%$ in 
\textsc{se2n}. These bodies will constitute the ``steeply 
infalling debris'' whose destiny was studied by \cite{Brown+17}. 
The remainder may undergo circularisation and collisional 
processing \citep[e.g.,][]{Veras+15,KenyonBromley17}. 
Before these processes occur, however, the bodies 
may be ejected, particularly when the 
planets are massive. In the \textsc{s2sj} runs, $82\%$ of 
bodies which crossed the Roche limit were subsequently 
ejected from the system, with a median lifetime to ejection 
of only $18$\,Myr. As planet masses are reduced, ejection 
becomes less efficient and takes longer: in \textsc{n2s}, 
$65\%$ of these bodies are ejected, with a median 
lifetime of $152$\,Myr, and in \textsc{se2n} 
only $43\%$, with a median lifetime of $677$\,Myr. 
Now, circularisation of debris streams from disrupted 
bodies may take many Myr \citep[e.g.,][]{Veras+15}. 
We might expect therefore that much of the disrupted 
material in the \textsc{s2sj} systems does not in fact make 
it down to the WD surface but will continue to be scattered 
by the planets and be ejected into interstellar space. On the 
other hand, most of the disrupted material in systems with 
low-mass planets will not be scattered out of the system 
but will make its way to the surface of the WD, whether 
through circularisation of the debris stream or ongoing 
gravitational interactions with the planets.

A final caveat here is that, for computational reasons, 
we did not include GR corrections in the numerical integrations. 
These will in general hinder the delivery of material 
to the WD, as they can suppress secular eccentricity forcing 
such as through the Kozai mechanism. The final delivery of 
material even by the low-mass planets may therefore be achieved 
not through purely dynamical forcing from the planets but by 
other forces acting to circularise the debris streams.

\subsubsection{Accretion at early WD cooling ages}

\begin{figure}
  \includegraphics[width=0.5\textwidth]{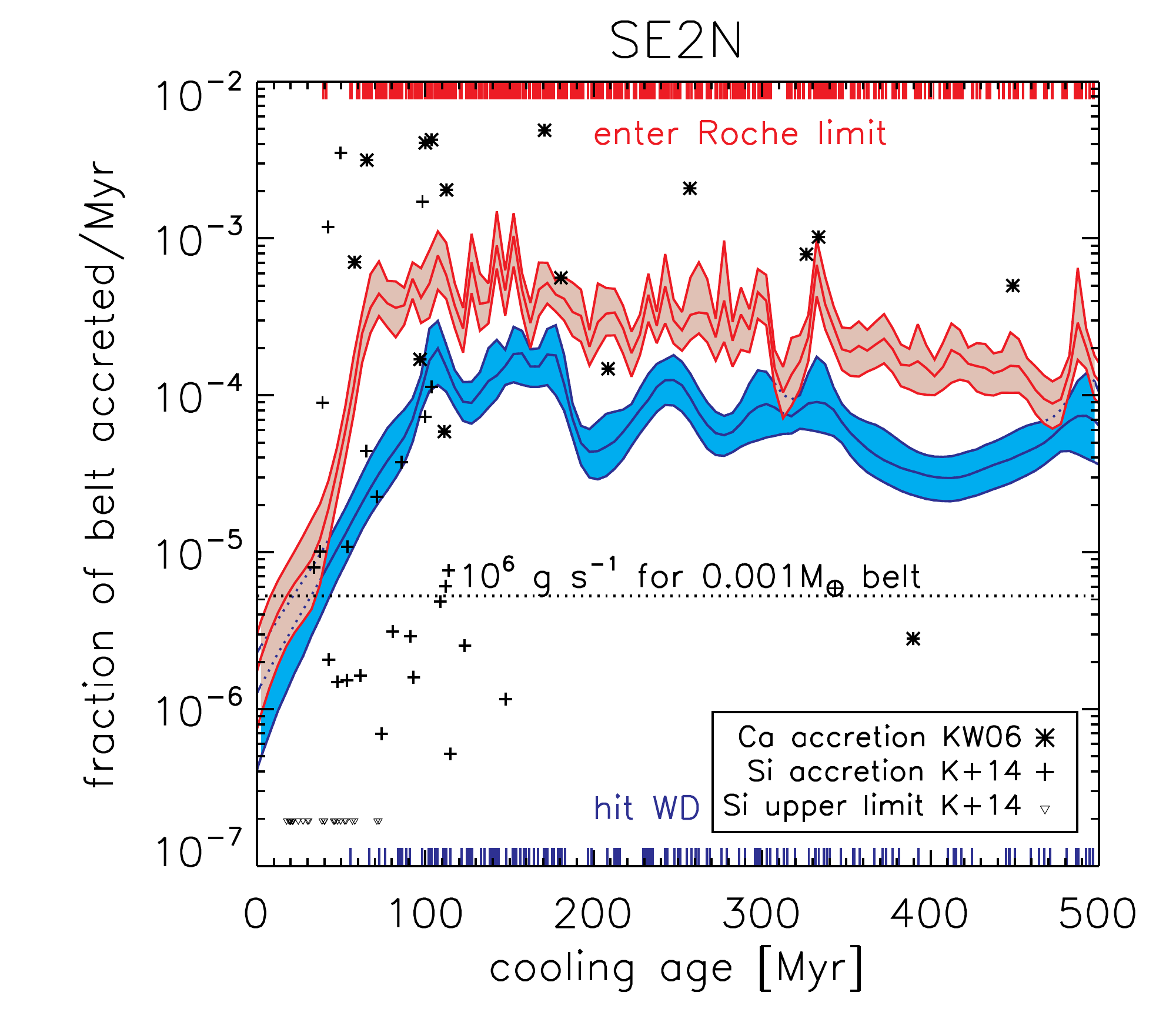}
  \caption{Accretion rates around young WDs. Here 
    both inner and outer belts from the \textsc{se2n} simulations 
    are combined, and the colours indicate the rates at which 
    particles cross the WD's Roche limit (red) or collide with 
    the WD (blue). No particles are delivered at young ages, 
    with particles beginning to cross the Roche limit at 
    $\sim40$\,Myr. This is approximately the age at which 
    accretion rates become detectable in the young DA sample 
    of \protect\cite{Koester+14}. The data from 
    \protect\cite{Koester+14} are shown as crosses 
    (Si-accreting systems) and triangles (upper limits), 
    and data from \protect\cite{KoesterWilken06} as 
    stars (Ca-accreting systems). Conversion of the observed 
    accretion rates to our $y$-axis assumes a belt mass 
    of $10^{-3}\mathrm{\,M}_\oplus$ and a conversion 
    of the elemental mass to the bulk Earth.}
  \label{fig:kde-zoom}
\end{figure}

Scattering following instabilities amongst low-mass 
planets well reproduces the trend in accretion at 
late times. We now consider accretion at young WD cooling 
ages of $\sim100$\,Myr. \cite{Koester+14} studied young 
DA WDs and found that accretion only became detectable at cooling ages 
of a few 10s of Myr. In Figure~\ref{fig:kde-zoom}
we show the inferred mass accretion rates from detections
of atmospheric Si \citep{Koester+14} and Ca \citep{KoesterWilken06}, 
where we assumed a modest belt mass of $10^{-3}\mathrm{\,M}_\oplus$.
Interestingly, the observational detections begin at exactly the 
ages at which delivery of material to the WD begins in our 
\textsc{se2n} simulations. The first particle to cross the 
Roche limit did so at a cooling age of 39\,Myr, and the 
first to strike the WD did so at a cooling age of 56\,Myr. 
Following this, the rate of delivery increased rapidly 
to a plateau, as shown in Figure~\ref{fig:kde-zoom}. 
The lack of accretion at young ages is partly 
due to the delay after stellar mass loss before 
orbit-crossing begins, and partly due to the 
delay in delivering material to the WD once the 
planetary instability begins: the earliest delivery 
of a particle to the WD in the \textsc{se2n} runs 
was 16\,Myr after the onset of orbit crossing. 
Thus, even if our planetary system had been slightly 
less widely-separated, so as to induce instabilities 
as soon as possible after stellar mass loss, 
there would still have been a delay in the 
delivery of material to the WD. 
Scattering amongst low-mass planets therefore correctly 
reproduces the trends in accretion rates onto young 
WDs as well as old. We caution that 
\cite{Koester+14} pointed out that the route from 
disruption to accretion may be different around 
these younger, hotter, WDs, which may result in 
accretion occurring in shorter, more intense bursts 
with a low duty cycle.

Surprisingly, we found almost no accretion in the 
systems prior to planetary orbit-crossing. 
Our simulations, integrating from the beginning of the MS 
through AGB mass loss to eventual planetary instability, 
naturally include the growth of chaotic regions associated 
with mean-motion resonances with the planets, even before the 
planets begin significantly exciting each other's eccentricities as 
orbit-crossing begins.
We found that the growth of these chaotic regions, however, 
does not result in accretion onto the WD: across all simulations, 
one single particle crossed the WD's Roche limit (and 
was subsequently ejected) before planetary orbit-crossing 
began. This was located in the inner belt close to the 2:1 resonance
with the innermost planet in one of the \textsc{s2sj} runs, 
and survived a few 10s of Myr beyond
the formation of the WD before beginning to experience large eccentricity
excursions. This single particle represents a very low efficiency
of pollution ($0.09\%$ of the inner belt particles surviving the AGB).
We discuss this issue further in Section~\ref{sec:rates}.

\begin{figure}
  \includegraphics[width=0.5\textwidth]{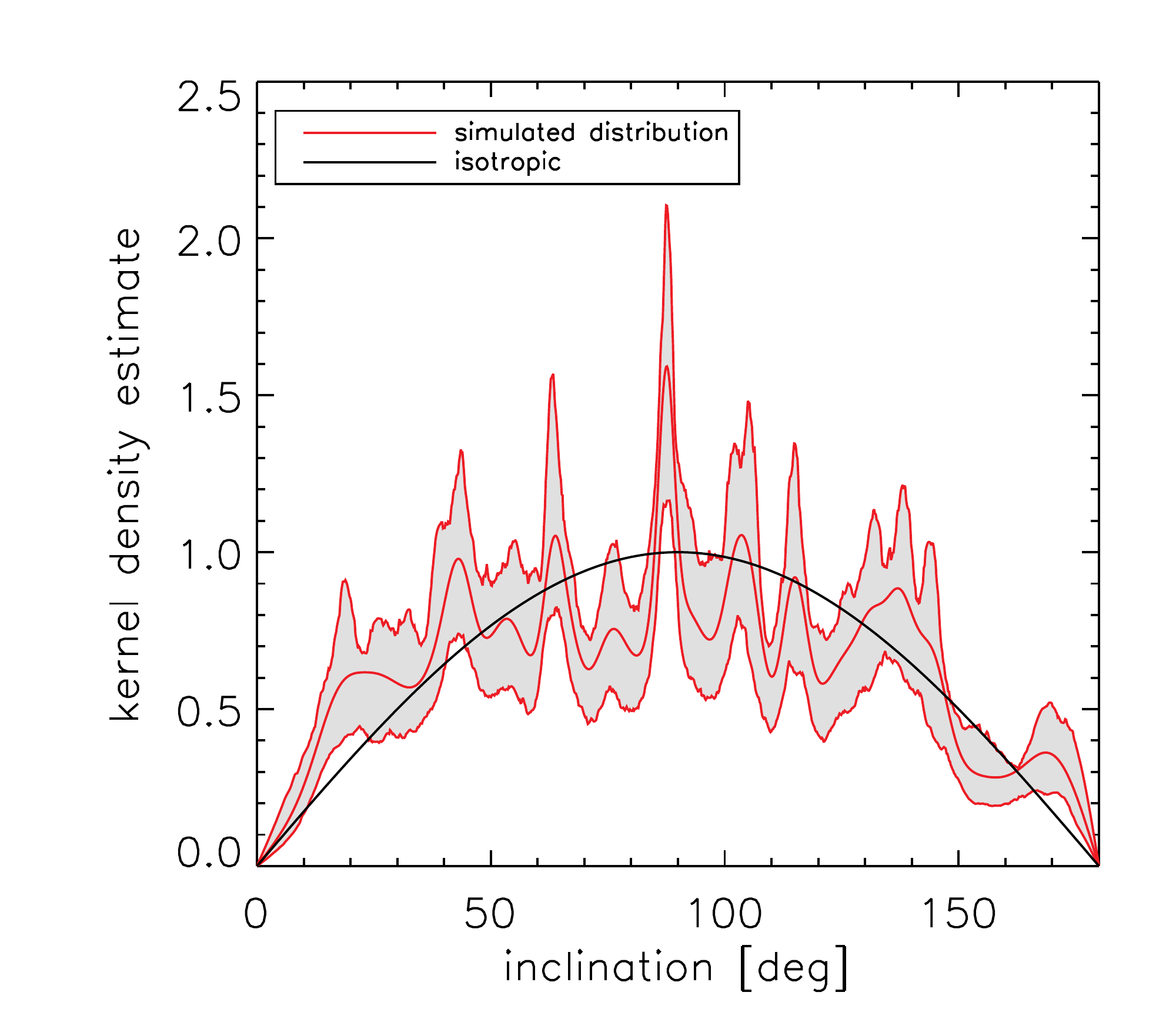}
  \includegraphics[width=0.5\textwidth]{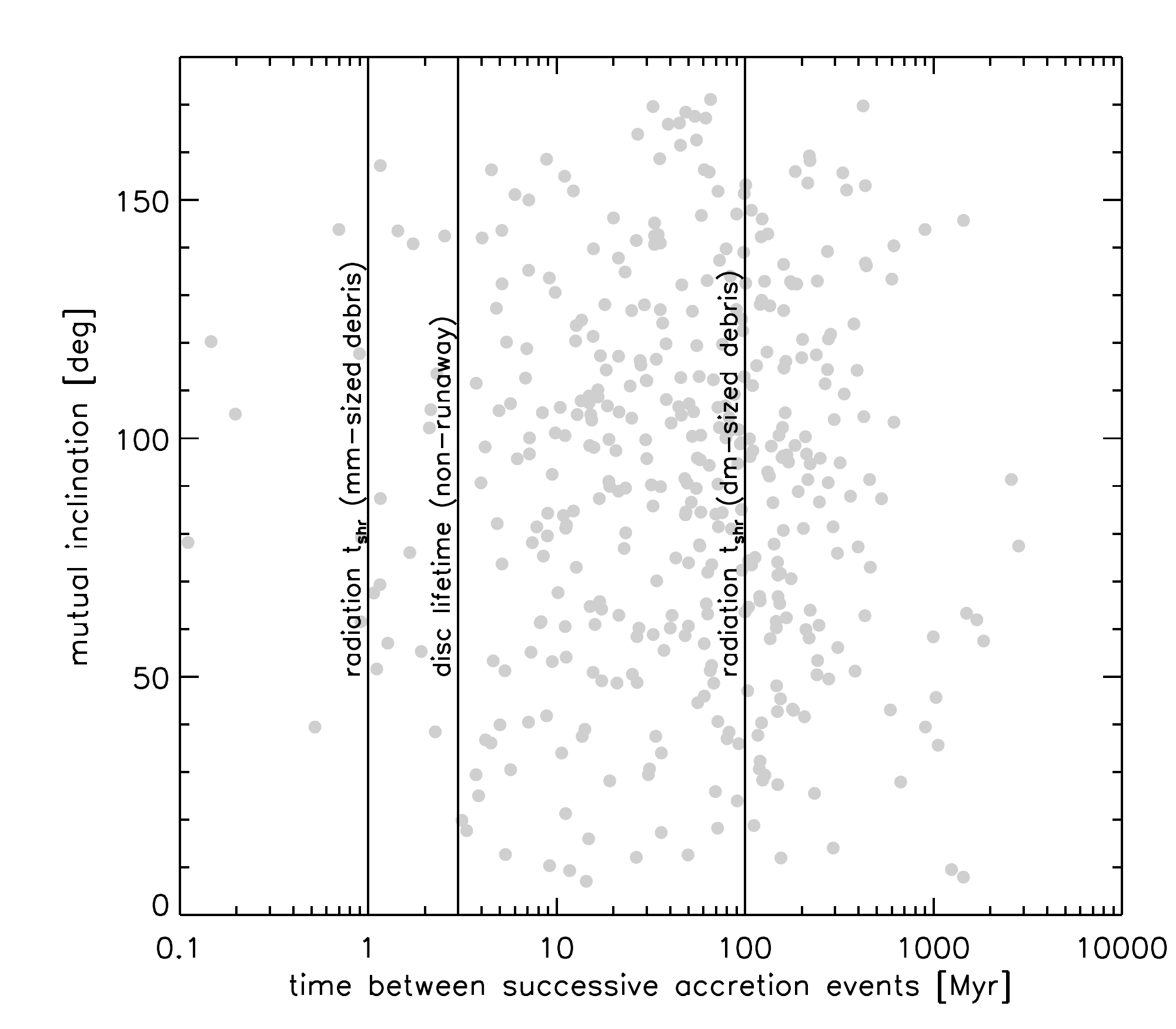}
  \caption{\textbf{Top: }Orbital inclinations, 
    with respect to the original reference plane, 
    of test particles removed upon colliding with the WD,
    in all the \textsc{se2n} runs. The distribution 
    is isotropic, with a weak excess at near-polar inclinations.
    \textbf{Bottom: }Intervals between successive accretion events 
    (here defined as striking the WD), and 
    mutual orbital inclinations of successive accreted planetesimals. The 
    distribution of these mutual inclinations is again isotropic, and no 
    correlation with the interval between events is seen. Vertical 
    lines show the approximate durations of two successive phases 
    leading to WD pollution: 
    the approximate lifetimes of debris belts circularising under 
    radiation forces \protect\citep[from][for two particle sizes]{Veras+15} 
    and the expected lifetime of circumstellar dust 
    discs \protect\citep[from][but there is considerable uncertainty 
      in these estimates, e.g., \citealt{Girven+12}]{Rafikov11}.}
  \label{fig:idist}
\end{figure}

\subsubsection{Orbital inclinations of bodies accreted onto the WD}

In our integrations we recorded the final position and 
velocity of objects colliding with the WD, 
allowing the instantaneous orbital inclination 
(with respect to the original reference plane of the planetary system) to be determined. 
Bodies approaching the WD come in with a roughly 
isotropic distribution of orbital inclinations (Fig~\ref{fig:idist}, top panel). 
This has several consequences. First, in systems exhibiting 
transiting asteroids such as WD~1145+017, we might not expect any 
correlation between the orbital inclination of the asteroids or 
disc and the inclinations of scattering bodies in the outer system. 
Second, we should not expect any correlation between the orbital 
planes of successively infalling bodies in any one system 
-- and indeed, the orbital planes of the asteroids of 
WD~1145+017 (edge-on) and the dust disc (not edge-on, as required 
to present a large enough emitting surface) probably have 
some mutual inclination \citep{Xu+18}. The lower panel of 
Fig~\ref{fig:idist} shows the time intervals between successive 
accretion events in the \textsc{se2n} runs, and the mutual 
inclinations of the orbits of successive infalling bodies. 
The mutual inclination distribution is again isotropic, and 
no correlation is seen with the time between accretion events, 
although all of our events in the \textsc{se2n} runs 
are separated by at least 0.1\,Myr\footnote{The \textsc{n2s} runs 
saw two events separated by $<1000$\,yr, 
with a mutual inclination of $160^\circ$.}. 
We also mark on some timescales relevant for later evolution 
of the disrupted planetesimals around the white dwarf: 
the timescale for orbit shrinking $t_\mathrm{shr}$ from 
\cite{Veras+15}, for debris of mm and dm sizes, and the 
lifetime of circumstellar dust discs (which are not undergoing 
the faster runaway accretion) from \cite{Rafikov11}. Note however that 
the disc lifetimes are highly uncertain: \cite{Metzger+12} estimated 
lifetimes of $1-10$\,Myr, similar to \citep{Rafikov11}, while semi-empirical 
analyses by \cite{Jura08} and \cite{Girven+12} argued for slightly 
shorter lifetimes of $0.01-1$\,Myr. 
Timescales of $\gtrsim1$\,Myr imply that in many cases the infalling 
planetesimals will meet tidal streams from previous disruption 
events in the process of circularisation, or circularised material 
now forming a disc of gas and dust round the WD 
\citep{Jura08,Wilson+14}. The high mutual 
inclinations mean that significant collisional processing, at 
very high velocities, will occur for the infalling material. 
Collisional damping at the high velocities in the neighbourhood 
of the WD is expected to be weak \citep{KenyonBromley17}, and 
so these high inclinations will be maintained as collisions occur. 
Significant collisional vapourisation of this material can therefore be expected.

\section{Discussion}

\label{sec:discuss}

\subsection{Current observations of prevalence and rates of accretion}

At least $25\%$ of WDs show spectroscopic evidence of 
metal pollution \citep[e.g.,][]{Koester+14}. If WD 
pollution is to be attributed largely to one dominant 
dynamical mechanism arising from one dominant system 
architecture, this architecture must therefore be found 
around a significant fraction of stars. In this sense, 
planets with masses in the super-Earth range are good 
candidates for being the perturbers that drive the 
accretion of material onto WDs. In the region probed 
by RV and transit surveys, within 1\,au around Solar-type 
stars, the planetary occurrence rate rises steeply 
towards smaller planets, and several 10s of per cent.\ of 
stars host at least one super-Earth or Neptune 
\citep{Mayor+11,Fressin+13}. This region will 
be totally cleared of planets by the star's 
RGB and AGB evolution 
\citep{VillaverLivio09,MustillVillaver12,Villaver+14}
However, if these occurrence rates 
also apply to planets at $\gtrsim5$\,au around 
intermediate-mass stars, low-mass planets are immediately more 
promising as a driver of pollution than are 
systems of unstable gas giants. 

The main weakness here is that for pollution 
to occur the planetary orbits themselves 
are required to be destabilised by AGB mass loss: recall 
that across all simulations we found only one single 
planetesimal that crossed the WD's Roche limit before 
planetary orbit-crossing began. As we showed in our 
preliminary simulations, the systems of super-Earths 
must be separated by $\sim6-10$ mutual Hill radii 
to favour stability on the Main Sequence followed 
by instability after the star becomes a WD 
(Table~\ref{tab:notp} and Figure~\ref{fig:delta-delta}). 
If we continue our use of known close-in systems as 
an analogy for the wider systems orbiting WD progenitors, 
we must note that the typical multi-planet system 
discovered by \emph{Kepler} has a separation of $\sim10-30$ 
mutual Hill radii \citep{Weiss+17}, which would imply that 
most systems probably remain stable following AGB mass loss. 
Alternatively, one could be optimistic and predict that 
the planetary systems orbiting intermediate-mass stars 
have a tighter spacing than those found on close orbits 
around Solar-type stars. From Table~\ref{tab:notp}, we 
see that a pollution occurrence rate of 
$25\%$ could be attained if all intermediate-mass stars 
have multiple super-Earths at $\sim10$\,au, and they 
are separated by $\sim10$ mutual Hill radii.

The time dependence of the accretion rates is an 
important test for evaluating mechanisms of WD pollution. 
Hollands et al (submitted) have recently measured WD accretion rates 
in a sample of old DZ WDs, with cooling ages from 1 to 8\,Gyr, allowing a 
comparison of our simulation results with an observational sample 
spanning a large range of cooling ages. They find an exponential 
decay in the accretion rates with age, with an e-folding time 
of $0.95$\,Gyr. This trend is overplotted in the upper panels of 
Figure~\ref{fig:tengulf} and the lower panel of Figure~\ref{fig:f_acc}. 
A very good match to the \textsc{se2n} 
simulations is seen, particularly those with outer planetesimal 
belts. Those with inner belts show a decay that is a little 
steep at late times, although this discrepancy disappears if we use 
passage of the Roche limit, rather than collision with the WD, 
as a criterion for accretion (Figure~\ref{fig:f_acc}). 
In contrast, the more massive planets in 
simulations \textsc{n2s} and \textsc{s2sj} both result in 
bursts of accretion that end very soon after instability. 
Moreover, very few planetesimals are left following 
instability among high-mass planets, so such systems experiencing 
an instability will not have a reservoir of planetesimals 
to draw on should they also host lower mass planets. 
This means that the observed pattern of WD pollution rates 
is inconsistent with large numbers of intermediate-mass stars hosting 
unstable systems of giant planets 
\citep[consistent with direct imaging surveys of A-type stars;][]{Vigan+17}, 
but it is consistent with 
large numbers hosting unstable systems of super-Earths. 
As we discussed above, this is believable in light of 
planetary occurrence rates around Solar-type stars. 

The observed time dependence of accretion rates onto 
young WDs is a second important constraint. \cite{Koester+14} 
found no detectable accretion onto DA WDs younger than 
a few 10s of Myr, after which the accretion was detected. 
This was not an observational bias as the upper limits 
for the younger WDs are below the detected rates around 
the older WDs. This is a second observational trend that 
our \textsc{se2n} simulations successfully reproduce 
(Figure~\ref{fig:kde-zoom}). 
The first particle to cross the Roche limit in our simulations 
did so at a cooling age of $39$\,Myr, while the 
first to directly collide with the WD did so at $56$\,Myr. 
After this, accretion rates quickly reach a plateau 
and then begin the slow decay discussed above.

The final consideration is the absolute scaling of 
the accretion rates. The accretion rates inferred 
from spectroscopic observations of WDs range from 
$\sim10^5-10^9\mathrm{\,g\,s}^{-1}$ 
\citep[e.g.,][]{Koester+14}, or 
$\sim5\times10^{-10}-5\times10^{-6}
\mathrm{\,M}_\oplus\mathrm{\,Myr}^{-1}$.
Let us take $10^9\mathrm{\,g\,s}^{-1}$ as a high 
accretion rate. If this were to be maintained in a system 
for 5\,Gyr, it would imply a total of $0.026\mathrm{\,M_\oplus}$ 
of material delivered in total in this time. If low-mass planets 
deliver material to the WD with $\sim20\%$ efficiency, this 
implies a total belt mass of $\sim0.13\mathrm{\,M}_\oplus$. 
This can be compared with the Solar System's Asteroid Belt 
mass of $5\times10^{-4}\mathrm{\,M}_\oplus$ \citep{Pitjeva05}: 
at $20\%$ efficiency, our Asteroid Belt could maintain 
a delivery rate of $\sim2\times10^7\mathrm{\,g\,s}^{-1}$ over 
5\,Gyr. We note that these relatively low belt masses 
($\lesssim1\mathrm{\,M}_\oplus$) mean that our use of 
massless test particles to model the planetesimals 
is valid, since the dynamical friction they impose 
on the planets in the system will be negligible 
(see Appendix).

Higher accretion rates, if they are not transient, 
require higher belt masses, and there is a limit to 
the amount of material that can be present in a 
planetesimal belt set by collisional evolution: 
belts that are initially more massive experience 
faster collisional evolution and at late times 
the belt mass tends asymptotically to the same 
level, regardless of the initial mass \citep{Wyatt+07}. 
For $3\mathrm{\,M}_\odot$ progenitors, \cite{BonsorWyatt10} 
estimated that a belt centred at 10\,au and of 
width 5\,au should only preserve $10^{-4}\mathrm{\,M}_\oplus$ 
of material to the end of the AGB, although this 
estimate assumed a maximum planetesimal size of $2$\,km. The 
maximum permitted belt mass is linearly proportional to 
the maximum planetesimal size and so belts containing larger 
planetesimals of a few hundred km in radius should be 
able to support the highest accretion rates of 
$\sim10^9\mathrm{\,g\,s}^{-1}$.

More distant belts experience much slower collisional 
evolution: for a fixed ratio of belt width $\Delta r$ 
to radius $r$, the collision timescale and maximum 
belt mass scale as $r^{13/3}$. Despite the lower 
efficiency of delivery from the more distant outer belts, 
the increased mass reservoir they support means that 
they can support higher accretion rates. However, recall 
the compositional constraint introduced in Section~\ref{sec:intro}: 
few polluted WDs show evidence for significant quantities 
of water. This implies that most source material originates 
relatively close to the star. Recent work by \cite{MalamudPerets17a} 
has found that bodies of 100\,km in size orbiting initially at 10\,au 
around a $3\mathrm{\,M}_\odot$ progenitor will be depleted of 
water on the AGB, while larger or more distant bodies 
will retain some fraction \citep{MalamudPerets17b}.

This implies that the inner belts 
in our simulations should be entirely water-depleted except 
for the largest, rarest bodies, and so our inner belt simulations 
match the observational constraints from the point of view of 
composition as well as the rate of accretion and its time 
dependence.

\subsection{Comparison of rates from this paper to other proposed mechanisms}

\label{sec:rates}

We now compare the accretion rates determined in this paper 
to those calculated in other scenarios. \cite{FrewenHansen14} 
found very high accretion efficiencies for a single 
eccentric planet embedded in a planetesimal belt, approaching 100\% of 
\emph{unstable} belt particles being accreted onto the star 
for low-mass ($10\mathrm{\,M}_\oplus$), eccentric ($e=0.8$) planets, 
at a rate of $7\times10^{-5}$ of the unstable belt mass per Myr 
at a cooling age of 1\,Gyr. This fractional rate is comparable 
to the rates we attain at a similar time in our \textsc{se2n} 
simulations. As we argued above, this is sufficient to reproduce 
the observed accretion rates. The advantages our model have over that 
of \cite{FrewenHansen14} are: self-consistent dynamical modelling 
to explain the presence of such an eccentric planet in a disc 
of planetesimals; the ability to destabilise a wider annulus of 
planetesimals, meaning a greater mass reservoir; and a larger 
semi-major axis, allowing more mass to survive collisional 
grinding on the Main Sequence.

The pre-instability phase of our simulations encompasses 
two related phenomena studied as potential mechanisms 
of WD pollution in previous works: the expansion of 
the chaotic zone of resonance overlap surrounding a 
planet's orbit \citep{Wisdom80,Bonsor+11,MustillWyatt12} 
and the expansion of individual unstable mean motion resonances 
\citep{Debes+12}. \cite{Bonsor+11} studied an outer belt 
surrounding a single planet on a circular orbit. They 
were able to obtain high accretion rates of 
$10^9-10^{10}\mathrm{\,g\,s}^{-1}$, but only on the 
assumption that material could be efficiently transported 
to the WD after scattering: with a single planet, 
particles did not attain pericentres less than $0.3$ times 
the planetary semimajor axis. Further transport requires 
extra planets in the system. We did not find any particles 
transported to the Roche limit from our outer belts 
in our three-planet simulations prior to planetary instability. 
This is not surprising, since further work by 
\cite{Bonsor+12} found that higher-multiplicity systems 
are required to deliver material even to $\sim1$\,au. 

It is perhaps more surprising that we did not see 
a significant contribution from the 2:1 interior MMR.
\cite{Debes+12} showed that, in the case of the Solar System, 
this unstable resonance with Jupiter will broaden following 
AGB mass loss and provide a source of material to pollute 
the WD, with $\sim2\%$ of their modelled asteroids 
crossing the WD's Roche limit. In contrast, we found 
only one such disrupted body, in the \textsc{s2sj} runs, 
accounting for only $0.9\%$ of that belt, or 
$0.1\%$ over all 10 simulations of inner belts. 
A large amount of this discrepency can be attributed to 
our distribution of particles, which was much broader 
($3.5$\,au in the \textsc{s2sj-inner} runs) than 
the source region around the 2:1 resonance most relevant 
for the delivery mechanism of \cite{Debes+12} (which would be $0.6$\,au 
for our belt radius). Thus, only 17\% of our belt is 
located around the 2:1 MMR. If we apply the 2\% survival 
rate from \cite{Debes+12} to 17\% of the belt particles 
surviving to the AGB tip (1069 across all \textsc{s2sj-inner} 
runs), we expect $3.6$ particles to enter the WD Roche limit 
prior to planetary instability. This is consistent with our single 
disruption (the probability of seeing 0 or 1 disruptions with a 
Poisson rate of $3.6$ is $12.6\%$).

With such small number statistics, we cannot say 
definitively that our simulations are less efficient at delivery 
from the 2:1 MMR. However, there are reasons to suspect that 
they should be. Recall that, in 
the Solar System, the unstable resonances are cleared by 
secular resonances and by secondary resonances between 
the frequency of the mean motion resonance itself and other 
frequencies in the system, with the 2:1 being cleared by 
a secondary resonance between the resonant libration period 
and the apsidal period whose efficiency depends upon 
the eccentricities and masses of Jupiter and Saturn \citep{Lecar+01}. 
Perhaps our Solar System's Kirkwood Gaps are unusually efficient 
at delivering material to the Roche limit, or the systems 
we integrated in this paper unusually inefficient. In particular, 
the eccentricities of the planets prior to instability are about an 
order of magnitude lower in our simulations than the eccentricities 
of Jupiter and Neptune. This has a strong effect on the stability 
of orbits in Jupiter's 2:1 MMR \citep[see Figure~7 of][]{Lecar+01}.

A final possibility is that the difference in disruption rates 
is a artefact of the choice of integrator used. We have used RADAU 
for the integrations in this paper, while \cite{Debes+12} used 
Bulirsch--Stoer. We also ran the \textsc{s2sj-inner} simulations with 
a Bulirsch--Stoer integrator, finding two particles crossing the WD's 
Roche limit prior to planetary orbit-crossing. Given the small number statistics, 
we do not have enough evidence to claim a systematic difference between the RADAU 
and BS integrators in this specific case.

\section{Conclusions}

\label{sec:conclude}

\begin{itemize}
\item We show, in self-consistent $N$-body integrations 
  lasting throughout a star's life, that planet--planet scattering 
  in the presence of a planetesimal disc can provide ongoing 
  accretion onto the WD for several Gyr.
\item Planet--planet scattering triggered by 
  post-main sequence stellar mass loss provides 
  a natural way of generating the eccentric planetary orbits 
  that are optimal for efficient delivery of planetesimals 
  to the WD. Following scattering, 
  planetary eccentricities remain high, since 
  the planetesimal disc masses required to explain 
  observed accretion rates are not high enough for 
  significant dynamical friction.
\item Low-mass planets (super-Earth to Neptune)
  are highly efficient at delivering material, and this
  delivery is ongoing, albeit at a decreasing rate, for
  several Gyr (Figure~\ref{fig:tengulf}). The decay rate is
  similar to that recently measured in DZ WDs
  (Hollands et al submitted).
\item Low-mass planetary systems also reproduce the observed delay
  of a few 10s of Myr before the onset of accretion onto a
  newborn white dwarf \citep[][and Figure~\ref{fig:kde-zoom}]
  {Koester+14}.
\item Instabilities in systems of Jovian planets 
  result in little accretion, and this is confined to 
  a narrow window of a few 10s of Myr.
\item Low-mass gas giants (Neptune to Saturn) deliver 
  a larger fraction of the belt to the WD than do higher 
  mass giants, but pollution 
  remains confined to a brief time window of a few 
  hundred Myr.
\item The ability of the scattering planets to 
  sweep up material from a relatively wide belt 
  ensures a large mass reservoir that can 
  provide sufficient material to explain the observed 
  accretion rates.
\item Material accreted onto the WD has a roughly 
  isotropic distribtion of orbital inclinations 
  (Figure~\ref{fig:idist}). 
  Thus, scattered asteroids are likely to meet other 
  asteroids, debris streams or discs at high mutual 
  inclinations. Significant collisional 
  processing---fragmentation and vapourisation---of 
  material may therefore take place before it is 
  finally accreted onto the WD.
\item The main challenge to our model comes from the unknown distribution of 
  planetary separations prior to planetary instability. We find that 
  planets separated by $6-10$ mutual Hill radii on the main sequence 
  will mostly survive main sequence evolution and become unstable after 
  mass loss (Table~\ref{tab:notp}, Figure~\ref{fig:delta-delta}). If 
  the wide-orbit low-mass planets we require are more widely separated, 
  however, the incidence of planetary instability may be insufficient 
  to explain the prevalence of WD pollution. On the other hand, if 
  planet--planet scattering is indeed responsible for the pollution 
  of WDs, we predict that a large fraction $\gtrsim30\%$ of WD 
  progenitors will have close-packed systems of super-Earths 
  at presently undetectable separations of $\sim10$\,au.
\end{itemize}

\section*{Acknowledgements}

AJM is supported
by the project grant 2014.0017 ``IMPACT'' from 
the Knut \& Alice Wallenbergs Foundation. 
This work has benefited from computer hardware and 
travel funds from the Walter Gyllenbergs fund of the 
Royal Physiografical Society in Lund. 
EV acknowledges support from the Spanish Ministry
of Economy and Competitiveness (MINECO) under
grant AYA-2014-55840P.
BTG received funding from the European 
Research Council under the European Union's 
Seventh Framework Programme (FP/2007--2013)/ERC 
Grant Agreement n.~320964 (WDTracer). 
DV gratefully acknowledges the support of the STFC 
via an Ernest Rutherford Fellowship (grant ST/P003850/1). 
AB acknowledges the support of the Royal 
Society by a Dorothy Hodgkin Fellowship. 
We thank Mark Hollands for providing the 
data on DZ pollution rates, and Jay Farihi, 
Jeremy Heyl and 
the anonymous referee for comments that improved 
the paper.

\appendix

\section{Integrator performance}

\label{sec:integrator}

\begin{figure}
  \includegraphics[width=0.5\textwidth]{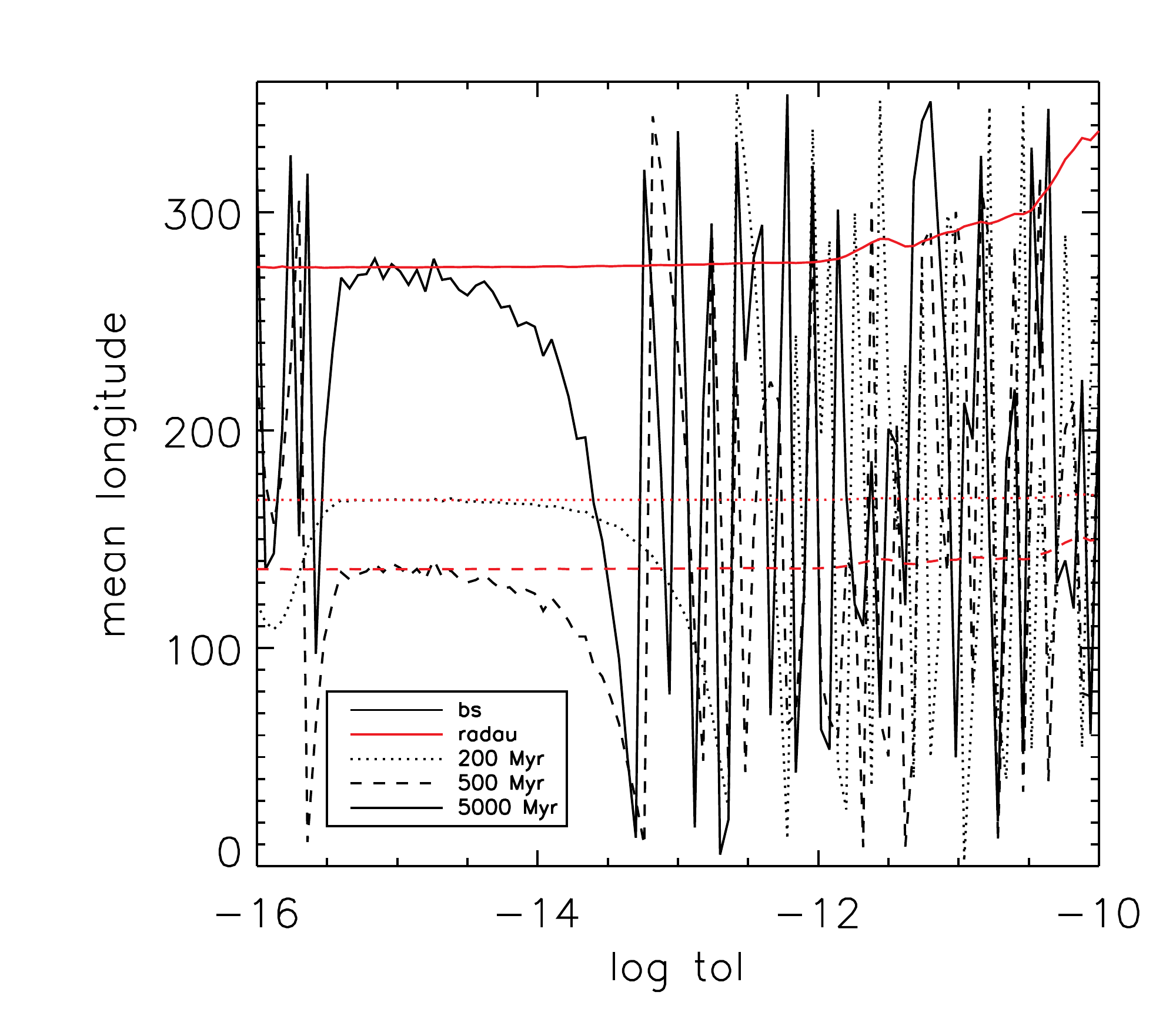}
  \caption{Performance of the RADAU integrator (adopted 
    for this paper) compared to the BS integrator. 
    We show the mean longitude of the planet of a single-planet 
    system as a function of integrator tolerance, at three times: 
    200\,Myr, 500\,Myr (just after mass loss) and 5\,Gyr. 
    RADAU exhibits excellent convergence, while BS cannot 
    accurately track the mean anomaly except in the narrow 
    range of tolerances around $10^{-15}$.}
  \label{fig:integrator}
\end{figure}

Here we briefly describe the integrator used 
for the simulations in this paper. A more complete 
description will be provided in Mustill et al (in prep.).

We adapt the RADAU integrator \citep{Everhart85} 
from the \textsc{Mercury} software package \citep{Chambers99}. 
Unlike the fixed-timestep hybrid symplectic integrator, 
RADAU can accurately handle eccentric orbits 
\citep[see, e.g., Figure~5 of][]{ReinSpiegel15}, a necessity 
for studying the delivery of material to the WD Roche 
limit or surface. Unlike BS, which uses a modified 
midpoint method, the RADAU integrator is a 
predictor--corrector method which utilises Gau{\ss}--Radau 
spacings. \cite{ReinSpiegel15} present an analysis 
of the performance of a (slightly modified) RADAU integrator 
compared to BS and symplectic algorithms, for the case of 
no mass loss. We modify the RADAU integrator to 
adjust the stellar mass and radius, fed in from 
an input file. As we showed with the 
BS integrator described in \cite{Veras+13}, 
it is essential to update the stellar mass not only 
at each major integrator timestep, but also at 
the subdivisions of these timesteps. This is done by 
linearly interpolating the mass and radius between the 
values recorded in the stellar evolution file.

We found that the modified RADAU integrator exhibits 
considerably better convergence properties than the BS 
integrator. In Figure~\ref{fig:integrator} we show the 
mean longitude of a planet initially at 10\,au 
in a single-planet system at three 
times into the integration: 200\,Myr (roughly halfway to 
the AGB tip), 500\,Myr (after the AGB tip) and 5000\,Myr 
(the end of the integration). These are plotted as 
a function of the stepwise error tolerance parameter. 
RADAU exhibits good convergence properties, with little 
difference in the final mean longitude for tolerance 
between $10^{-16}$ and $10^{-12}$, and only a small 
drift in final mean longitude for more lax tolerances. 
In contrast, BS only converges to the RADAU solution 
in a narrow tolerance range around $10^{-15}$, 
and the final mean longitude is essentially random 
for tolerances laxer than $10^{-13}$. RADAU is more 
computationally expensive than BS, and in this paper 
we adopt a tolerance of $10^{-11}$ as a compromise 
between speed and accuracy (this is $~25\%$ faster than 
running at a tolerance of $10^{-12}$). Although there is a small 
phase error by 5\,Gyr at this accuracy, it is negligible 
at 500\,Myr, a time more relevant for most of the 
dynamics.

\section{Importance of eccentricity damping}

\begin{figure}
  \includegraphics[width=0.5\textwidth]{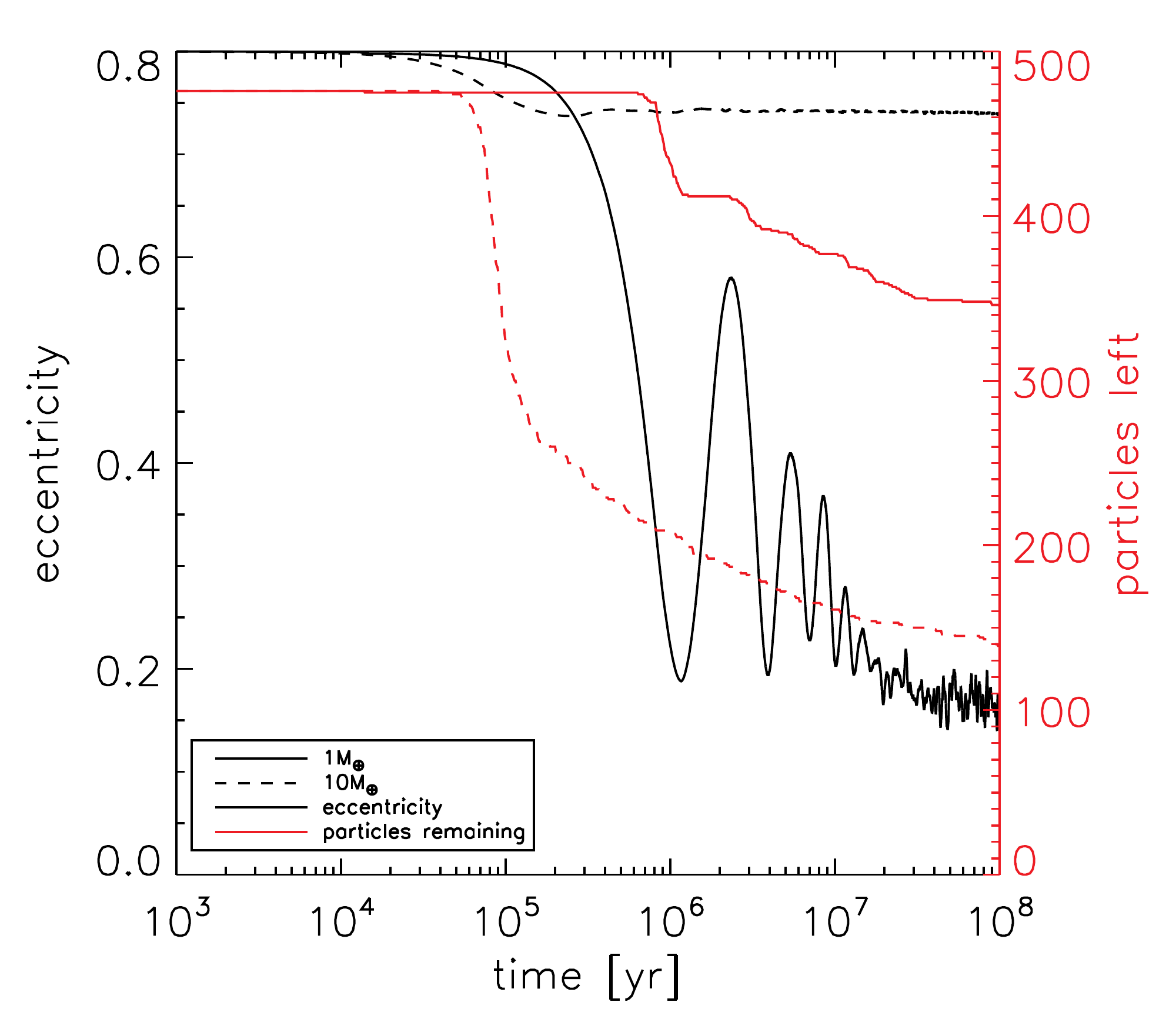}
  \caption{Systems of a single eccentric planet embedded in 
    a disc of massive test particles. The total disc mass is 
  $1\mathrm{\,M}_\oplus$. When the planet is of comparable mass, 
  strong eccentricity damping is seen, but only weak damping 
  for a planet ten times more massive.}
  \label{fig:1pl}
\end{figure}

We test the extent to which eccentricity damping may affect the ability of a highly 
eccentric planet to scatter particles. We consider a set-up similar to that 
of \cite{FrewenHansen14}, with a planet of eccentricity $e=0.8$ embedded in 
a disc of planetesimals on circular orbits. We take 476 planetesimals, with 
a total mass of $1\mathrm{\,M}_\oplus$. The planetesimals are massive test 
particles in \textsc{Mercury}: they interact gravitationally and collisionally 
with the planet and the star, but not with each other. 

The $10\mathrm{\,M}_\oplus$ planet's eccentricity falls approximately 10\% 
in the first 0.2\,Myr. In contrast, the $1\mathrm{\,M}_\oplus$ planet's 
eccentricity undergoes several damped secular cycles before stabilising at 
$~0.2$. This damping decouples it from the surviving disc particles, and 
far fewer are lost than with the higher-mass planet. 

It is therefore safe to ignore eccentricity damping when the disc is only 10\%
 of the planet's mass, but not for comparable masses.

\bibliographystyle{mnras}
\bibliography{pms8-3p+disc}

\bsp    
\label{lastpage}

\end{document}